\documentclass[11pt,a4paper]{article}
\usepackage{url}
\usepackage{jcappub}
\usepackage{amsfonts}
\usepackage{subfigure}
\usepackage{graphicx}
\usepackage{appendix}

\usepackage{epsfig}  
\usepackage{graphicx}   
\usepackage{slashed}       
\usepackage{tikz}
\usepackage{url}
\usepackage{color}
\usepackage{multirow}
\usepackage{comment}
\usepackage{amssymb}
\usepackage[capitalize]{cleveref}
\usepackage{bm}
\usepackage{soul}
\usepackage{orcidlink}
\usepackage[normalem]{ulem}

\DeclareMathOperator{\cm}{cm}
\DeclareMathOperator{\GeV}{GeV}
\DeclareMathOperator{\eV}{eV}

\DeclareMathOperator{\MeV}{MeV}

\DeclareMathOperator{\s}{s}

\DeclareMathOperator{\km}{km}

\DeclareMathOperator{\kpc}{kpc}
\DeclareMathOperator{\g}{g}

\DeclareUnicodeCharacter{2212}{-}

\allowdisplaybreaks

\bibpunct{[}{]}{,}{n}{}{,}
\definecolor{ForestGreen}{RGB}{34,139,34}

\newcommand{\fermi}{\textit{Fermi}}

\begin{document}

\title{Probing protoneutron stars with gamma-ray axionscopes}

\author[a,b]{Alessandro Lella~\orcidlink{0000-0002-3266-3154},}
\author[c,d]{Francesca Calore~\orcidlink{0000-0001-7722-6145},}
\author[f]{Pierluca Carenza~\orcidlink{0000-0002-8410-0345},}
\author[c,d,e]{Christopher Eckner~\orcidlink{0000-0002-5135-2909},}
\author[g,h]{Maurizio Giannotti~\orcidlink{0000-0001-9823-6262},}
\author[a,b,i,k]{Giuseppe Lucente~\orcidlink{0000-0002-3266-3154},}
\author[a,b]{Alessandro Mirizzi~\orcidlink{0000-0002-5382-3786}}

\affiliation[a]{Dipartimento Interateneo di Fisica ``Michelangelo Merlin'', Via Amendola 173, 70126 Bari, Italy}
\affiliation[b]{Istituto Nazionale di Fisica Nucleare - Sezione di Bari, Via Orabona 4, 70126 Bari, Italy}
\affiliation[c]{LAPTh, CNRS, F-74000 Annecy, France}
\affiliation[d]{LAPP, CNRS, F-74000 Annecy, France}
\affiliation[e]{University of Nova Gorica, Center for Astrophysics and Cosmology, Vipavska 11, SI-5000 Nova Gorica, Slovenia}
\affiliation[f]{The Oskar Klein Centre, Department of Physics, Stockholm University, Stockholm 106 91, Sweden}
\affiliation[g]{Physical Sciences, Barry University, 11300 NE 2nd Ave., Miami Shores, FL 33161, USA}
\affiliation[h]{Centro de Astropart{\'i}culas y F{\'i}sica de Altas Energ{\'i}as (CAPA), Universidad de Zaragoza, Zaragoza, 50009, Spain}
\affiliation[i]{Kirchhoff-Institut f\"ur Physik, Universit\"at Heidelberg, Im Neuenheimer Feld 227, 69120 Heidelberg, Germany}
\affiliation[k]{Institut f\"ur Theoretische Physik, Universit\"at Heidelberg, Philosophenweg 16, 69120 Heidelberg, Germany}

\emailAdd{alessandro.lella@ba.infn.it, calore@lapth.cnrs.fr, pierluca.carenza@fysik.su.se, eckner@lapth.cnrs.fr, mgiannotti@unizar.es, giuseppe.lucente@ba.infn.it, alessandro.mirizzi@ba.infn.it}

\abstract{
Axion-like particles (ALPs) coupled to nucleons can be efficiently produced in the interior of protoneutron stars (PNS) during supernova (SN) explosions. If these ALPs are also coupled to photons they can convert into gamma rays in the Galactic magnetic field. 
This SN-induced gamma-ray burst can be observable by gamma-ray telescopes like {\it Fermi}-LAT if the SN is in the field of view of the detector. 
We show that the observable gamma-ray spectrum is sensitive to the production processes in the SN core. 
In particular, if the nucleon-nucleon bremsstrahlung is the dominant axion production channel, one expects a thermal spectrum with average energy $E_a \simeq 50$ MeV. 
In this case the gamma-ray spectrum observation allows for the reconstruction of the PNS temperature. In case of a sizable pion abundance in the SN core, one expects a second spectral component peaked at $E_a\simeq 200$ MeV due to axion pionic processes. We demonstrate that, through a dedicated LAT analysis, we can detect the presence of this pionic contribution, showing that the detection of the spectral shape of the gamma-ray signal represents a unique probe of the pion abundance in the PNS.
}

\maketitle
\date{\today}

\section{Introduction}
\label{sec:intro}

Core-collapse supernovae (SNe) are unique astrophysical laboratories  for weakly-interacting particles such as axions 
and axion-like particles (ALPs) (see Ref.~\cite{Caputo:2024oqc} for a recent review). 
Because of the extreme conditions of temperature and density expected in the inner regions of the protoneutron star (PNS), ALPs could be copiously produced exploiting their coupling to nucleons 
$g_{aN}$, with $N=n,\,p$ for neutrons and protons respectively~\cite{Lella:2022uwi,Lella:2023bfb}. \emph{Indirect} constraints on this flux can be obtained from the impact of the ALP emissivity on the duration of the SN neutrino burst. Notably, the SN 1987A observation allows one to constrain $g_{aN} \lesssim 10^{-9}$~\cite{Lella:2023bfb}. 
In case ALPs are also coupled to photons, the SN ALP flux can convert into gamma rays in the Galactic magnetic field~\cite{Payez:2014xsa},
opening the possibility for a \emph{direct} detection of a SN ALP-induced gamma-ray burst in
gamma-ray telescopes~\cite{Meyer:2016wrm,Calore:2023srn}. In this regard, at the time of  SN 1987A explosion the non-observation of gamma rays in the Solar 
Maximum Mission Satellite~\cite{Payez:2014xsa} allows one to exclude values of the ALP-photon coupling 
$g_{a\gamma} \gtrsim 10^{-15}$ GeV$^{-1}$ for ALP masses
 $m_a <10^{-10}$ eV~\cite{Calore:2020tjw}. 

In this work, we explore the physics potential of a possible detection,
in a gamma-ray telescope like the {\it Fermi} Large Area Telescope ({\it Fermi}-LAT),
of an ALP flux produced 
by a Galactic SN via the ALP-nucleon coupling and  converting into gamma rays  in the Milky-Way magnetic field thanks to the ALP-photon coupling $g_{a\gamma}$~\cite{Meyer:2016wrm, Calore:2023srn,Muller:2023pip}. 
We will show how the detection of a gamma-ray burst can provide valuable information to probe the core of a SN, complementary to the ones obtained by SN neutrino observations in large underground detectors (see, e.g. Ref.~\cite{Mirizzi:2015eza} for a review). In particular, the detection of the neutrino burst from the next nearby SN will help the entire community to shed light over the SN explosion mechanism. However, since neutrinos are emitted from the edge of the PNS, they will be not sensitive to the main conditions characterizing the interior of the SN core, where neutrinos are trapped. On the other hand, ALPs (or other FIPs) can be copiously produced in these regions, and, for the values of the ALP couplings considered in this work, they can freely escape the inner PNS. As a consequence, the detection of an axion burst could provide us with unique information about the inner SN core.

In this context, in recent years the characterization of ALP production in nuclear processes has revealed to be more complex than originally thought in some of the pioneering works of the '80s and '90s~\cite{Carena:1988kr,Brinkmann:1988vi,Raffelt:1993ix,Raffelt:1996wa,Turner:1991ax,Keil:1996ju}. 
At first, the dominant ALP production channel in a hot and dense nuclear medium  was considered to be nucleon-nucleon ($NN$) bremsstrahlung~\cite{Carena:1988kr,Brinkmann:1988vi,Turner:1991ax,Raffelt:1993ix,Raffelt:1996wa,Carenza:2019pxu}.
In this case one would expect a thermal energy spectrum for SN ALPs, with average energy 
$E_a\sim 50$ MeV. We will show that, in this case, the detection of the SN ALP-induced gamma-ray signal would allow one to reconstruct the average temperature of the SN core and its time evolution in the first seconds after the core bounce (see Refs.~\cite{OHare:2020wum, Hoof:2023jol} for recent works using ALPs to probe the solar magnetic field and temperature profiles).
Remarkably, ALPs can probe a complementary region with respect to neutrinos, whose emission occurs from a thin layer around the \emph{neutrino-sphere} at $R_\nu\simeq 30\,\km$~\cite{Mirizzi:2015eza}.\\
The ALP production only through $NN$ bremsstrahlung was questioned after the seminal contribution in Ref.~\cite{Fore:2019wib}, which re-estimated the abundance of negatively charged pions in the SN core beyond the ideal gas approximation. The behavior of pionic matter inside the hot and dense PNS is currently an extremely relevant topic, still under investigation (see Ref.~\cite{Fore:2023gwv} for recent developments). In particular, the role of pions inside SN simulations has been overlooked for a long time, since they are expected not to give a significant contribution to the explosion mechanism. 
Nevertheless, in Ref.~\cite{Fore:2019wib} the authors proposed that strong interactions in the PNS can enhance the fraction of negatively charged pions. The presence of a large fraction of negative pions (1 to 3 \% of the nucleon fraction) induces a competitive ALP production mechanism from the Compton scattering of pions on nucleons~\cite{Keil:1996ju,Carenza:2020cis} ($\pi N$ hereafter), which in several conditions is expected to dominate~\cite{Carenza:2020cis}.
In this case one would expect a second spectral component to peak at higher energies, i.e.~$E_a\sim 200$ MeV due to ALP pionic processes. As we shall discuss, the {\it Fermi}-LAT observation of the gamma-ray flux may help to reveal the relative importance of the bremsstrahlung and pionic processes in the SN environment. Thus, the detection of the peak associated with pion conversions can discriminate whether or not pions are present inside the SN core. Conversely, the pion fraction in the SN core is expected to have little impact on the observable neutrino signal. Therefore, the SN axion signal would represent a unique opportunity to answer the question related to the pion abundance in the PNS, since there is no other messenger sensitive to this feature.

Our work is structured as follows.
In Sec.~\ref{sec:ALP production}, we revisit the ALP production in a SN nuclear medium due to $NN$ and $\pi N$ processes, we provide simple fitting expressions for the ALP energy spectra, and we show relations between the PNS temperature and the average energy of the ALP spectrum. In Sec.~\ref{sec:Conversions} we characterize the ALP-photon conversions in the Milky-Way magnetic field. In Sec.~\ref{sec:FermiLAT-analysis} we present our analysis of the gamma-ray energy spectrum observable in \fermi-LAT. In Sec.~\ref{sec:Results} we discuss the reconstruction of the fitting parameters in \fermi-LAT. In particular, we show how it is possible to infer information on the pion abundance in the SN core and to reconstruct the time evolution of the average core temperature.
Finally, in Sec.~\ref{sec:Conclusion} we summarize our results and we conclude.
The main text is followed by Appendix~\ref{app:Parameters}, in which we report all the fitting parameters and fitting formulae employed for the ALP spectrum.

\section{ALP production in the PNS nuclear medium}
\label{sec:ALP production}
\subsection{Production processes}

The ALP production in the nuclear medium of the PNS is due to $NN$ bremsstrahlung $N+N\rightarrow N+N+a$ and $\pi N$ pionic Compton-like processes  $\pi+N \rightarrow a+N$~\cite{Carena:1988kr,Brinkmann:1988vi,Turner:1991ax,Raffelt:1996wa,Keil:1996ju,Carenza:2019pxu,Carenza:2020cis,Lella:2022uwi,Lella:2023bfb,Carenza:2023lci}. The state-of-the-art calculation for the bremsstrahlung emission rate has been obtained in Ref.~\cite{Carenza:2019pxu} and accounts for corrections beyond the usual one-pion exchange (OPE) approximation, namely a non-vanishing mass for the exchanged pion~\cite{Stoica:2009zh}, the contribution from the two-pion exchange~\cite{Ericson:1988wr}, effective in-medium nucleon masses and multiple nucleon scatterings~\cite{Raffelt:1991pw,Janka:1995ir}. 
On the other hand, since the fraction of pions in the SN core was believed to be small, the contribution due to pionic processes has been overlooked for many years. Nevertheless, in Ref.~\cite{Fore:2019wib} the authors argued that strong interactions can enhance the density of negatively charged pions. Thus, the emission rate via pion conversions was re-estimated in Ref.~\cite{Carenza:2020cis}, realizing that it may be comparable and even dominant with respect to bremsstrahlung. Furthermore, the pion conversion emission rate computed in Ref.~\cite{Lella:2022uwi} took into account also the contribution due to the contact interaction term~\cite{Choi:2021ign} and the $\Delta$(1232) resonance~\cite{Ho:2022oaw}, which could even enhance the ALP emissivity. However, a self-consistent treatment of pionic matter in the PNS interior is not available yet, and, in SN models characterized by particularly high densities, pions may undergo Bose-Einstein condensation. Therefore, depending on our purpose, in the following we will consider both cases with and without the pion conversion contribution to ALP emissivities. In this work we will focus on ALPs in the \textit{free-streaming} regime $g_{aN}\lesssim10^{-8}$. In this regime, absorption effects by means of inverse nuclear processes can be neglected and ALPs can freely escape the PNS volume~\cite{Lella:2023bfb}.

\subsection{SN models}
\label{sec:SNmodels}
In this work we adopt as reference to calculate the SN ALP fluxes the 1D spherical symmetric {\tt GARCHING} group's SN model SFHo-s18.8 provided in Ref.~\cite{SNarchive} and based on the neutrino-hydrodynamics code 
{\tt PROMETHEUS-VERTEX}~\cite{Rampp:2002bq}. The simulation employs the SFHo Equation of State (EoS)~\cite{Hempel:2009mc,Steiner:2012rk} and is launched from a stellar progenitor with mass $18.8~M_\odot$~\cite{Sukhbold:2017cnt}, leading to neutron star (NS) with baryonic mass $1.35~M_\odot$.\\
We highlight that we have accounted for the presence of pions in the SN core by following the procedure illustrated in Ref.~\cite{Fischer:2021jfm}, including the pion-nucleon interaction as described in Ref.~\cite{Fore:2019wib}. Indeed, due to strong interactions pions are in chemical equilibrium in the SN core. Hence, their abundances can be calculated from the corresponding local thermal and chemical equilibrium conditions for a gas of massive and relativistic bosons. This implies that pions have a chemical potential $\mu_{\pi^{\pm}}=\mp \hat{\mu}$, where $\hat{\mu} = \mu_n - \mu_p$ is the difference between the neutron and the proton chemical potential, and $\mu_{\pi^0}=0$~\cite{Fischer:2021jfm}. Therefore, due to the Boltzmann suppression factor, the abundance of negative pions $Y_{\pi^-}\propto \exp(\hat{\mu}/T)$ in the SN core is enhanced, compared to $\pi^0$ and $\pi^+$, when $\hat{\mu}\simeq m_\pi \simeq 139~\MeV$~\cite{Carenza:2020cis,Fischer:2021jfm}. For $\hat{\mu}>m_{\pi}$ a Bose-Einstein condensate of pions is favored, but its formation at SN densities is under debate~\cite{Migdal:1990vm}. Since our reference model does not include pions, we compute the pion abundance and chemical potential used in the calculation of the ALP emissivity (see Refs.~\cite{Carenza:2020cis,Lella:2022uwi,Lella:2023bfb} for more details) by imposing $\mu_{\pi^{-}}=\hat{\mu}$, where $\hat{\mu}$ is provided by the SN model.
This post-processing addition of pions is justified as long as the impact of pionic matter on the PNS properties is not larger than the impact of muons. Nevertheless, since the density in the inner PNS increases during the cooling phase, it is possible to observe the formation of a Bose-Einstein condensate of pions when $\hat{\mu}>m_{\pi}$.
In particular, our benchmark SN model shows remarkable effects due to pion condensation starting from post-bounce times $t_{\rm pb}\gtrsim3\s$. Since the emission of ALPs from a pionic condensate has not been investigated yet, we adopt in the following the most conservative approach, by setting pion-conversion processes to zero in all the regions in which we observe pion condensation.\\
To test the validity of the results reported in the following Sections on different SN simulations~[see, e.g., Sec.~\ref{sec:TemperatureReconstruction}], it is convenient to introduce two other different models. The first one, labeled in literature as {SFHo-s20}, is characterized by the same EoS as {SFHo-s18.8}, but it is launched from a $20\,M_\odot$ progenitor and leads to PNS baryonic mass $M_{\mathrm{NS}}\simeq1.95\,M_\odot$~\cite{Woosley:2007as}. The second one, labeled as LS220-s20, employs the LS220 EoS~\cite{Lattimer:1991nc}, characterized by a different behavior in the cooling phase with respect to SFHo-type simulations due to a different convective activity connected to the nuclear symmetry energy~\cite{Roberts:2011yw}. As for SFHo-s20, this model originates from a $20\,M_\odot$ progenitor and produces a NS with baryonic mass $M_{\mathrm{NS}}\simeq1.93\,M_\odot$~\cite{Woosley:2007as}.
Models characterized by high 
PNS masses, such as SFHo-s20 and LS220-s20, are typically characterized by values of temperatures and densities significantly higher than what observed for low PNS-mass models, such as SFHo-s18.8. Therefore, for SFHo-s20 and LS220-s20, the treatment for pionic matter described before may lead to the formation of a pionic condensate in the inner core since the very beginning of the SN cooling phase, at post-bounce times $t_{\rm pb}\gtrsim1\,\s$. For this reason, we opted for neglecting ALP production via pionic processes for the high PNS-mass SN models considered in this work.

\begin{figure} [t!]
\centering
    \includegraphics[width=1\columnwidth]{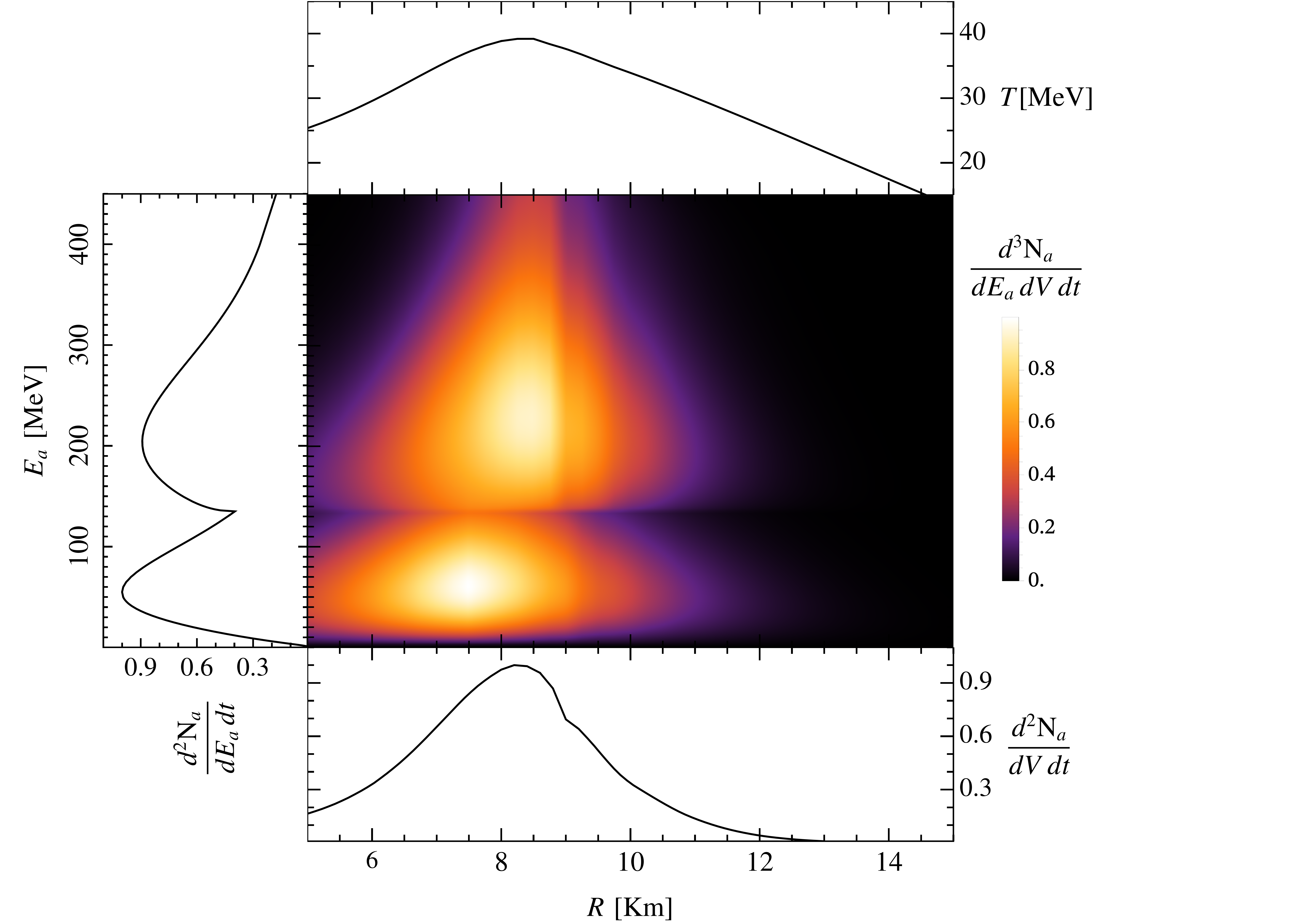}
    \caption{ALP production density (normalized at its peak value) for our benchmark SN model at $t_{\rm pb}=1\,\s$, displayed as a function of the PNS radius and of the energy of the produced ALP. \emph{Lower panel}: Energy-integrated ALP production density (normalized at its peak value) as a function of the production radius. \emph{Left panel}: Volume-integrated ALP production spectrum (normalized at its peak value). \emph{Upper panel}: Temperature profile of the SFHo-s18.8 model.}
	\label{fig:AxionProduction}
\end{figure}

\subsection{SN ALP spectra}
By evaluating the ALP spectra per unit time and unit volume $d^3N_a/dVdE_adt$~\cite{Lella:2022uwi} over the hydro-profiles associated to the SFHo-s18.8 model, we have obtained Fig.~\ref{fig:AxionProduction}. In particular, this figure refers to a single snapshot of our benchmark SN model at post-bounce time $t_{\rm pb}=1\,\s$. 
As we can deduce from the lower panel of Fig.~\ref{fig:AxionProduction}, ALP production is efficient in the region $6-10\,\km$, which is located around the temperature peak at $R\simeq8\,\km$. 
Remarkably, if pions are present in the SN core, the emission spectrum of free-streaming ALPs is characterized by a bimodal shape, since $NN$ bremsstrahlung and pionic processes are efficient~\cite{Carenza:2020cis,Lella:2022uwi,Lella:2023bfb} in different energy ranges. Indeed, the left panel of Fig.~\ref{fig:AxionProduction} shows that bremsstrahlung and pion conversion spectra peak at $E_a\sim 50\MeV$ and $E_a\sim 200 \MeV$, respectively. 
Notice that pionic production is maximum in the hottest regions of the core, since it is very sensitive to the SN temperature~(see Eq.~(5) of Ref.~\cite{Lella:2022uwi}). On the other hand, bremsstrahlung is more efficient at slightly smaller radii, where the nuclear density is higher.
These results confirm what suggested before, i.e. that the features of the ALP emission spectra encode many of the properties of the internal regions of the PNS, where ALPs are produced.

\begin{figure} [t!]
\centering
    \includegraphics[width=1\columnwidth]{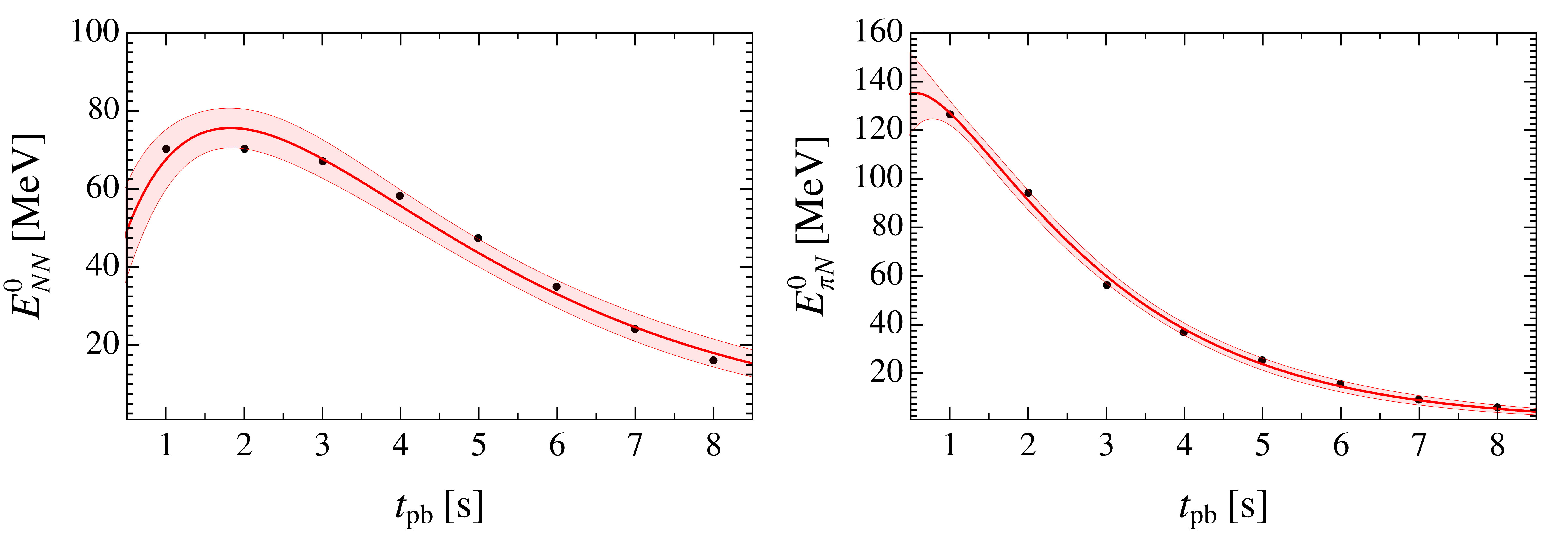}
    \includegraphics[width=1\columnwidth]{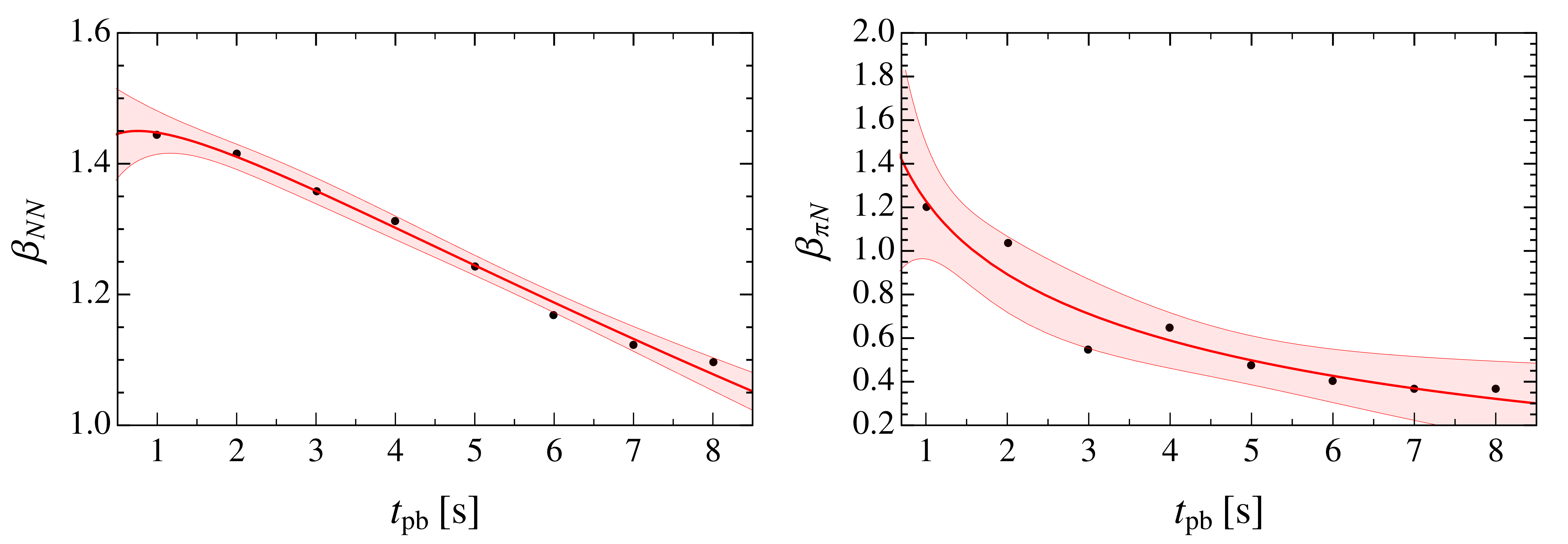}
    \includegraphics[width=1\columnwidth]{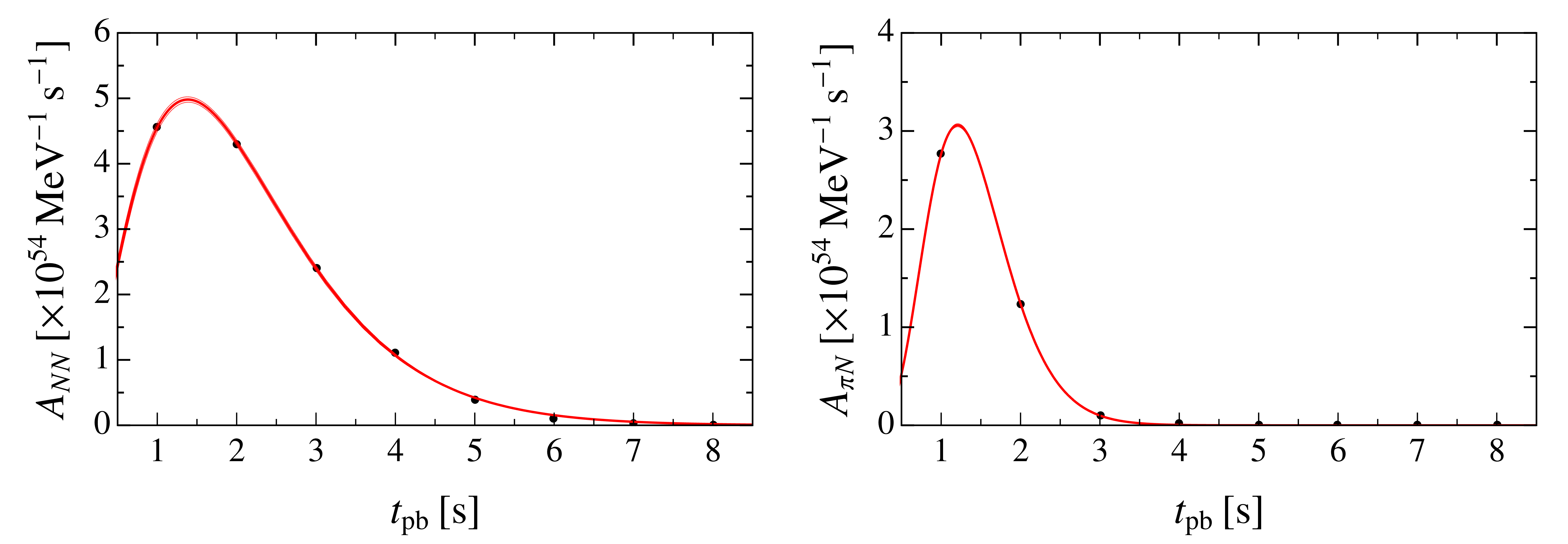}
    \caption{Time behavior of the fitting parameters for the SFHo-s18.8 model in the time interval $t_{\rm pb}\in [1,\,8]\,\s$. The red line displays the analytical fits for the considered parameters, whose explicit expressions are reported in Appendix~\ref{app:Parameters}. The red shadowed regions display 2-$\sigma$ uncertainty bands for the fit.}
	\label{fig:Parameters}
\end{figure}

\subsection{Fitting expressions for ALP spectra}
The complete expression for the ALP emission spectra by means of $NN$ bremsstrahlung and $\pi N$ pionic Compton-like processes is reported in Ref.~\cite{Lella:2022uwi}. However, since these expressions require complex numerical integration procedures,  in order to calculate the gamma-ray induced signal it is useful  to introduce some simple analytical fitting formulae. In analogy to Ref.~\cite{Lella:2022uwi}, the numerical fits will be obtained by assuming as a benchmark the ALP-neutron coupling $g_{an}=0$, inspired by the Kim-Shifman-Vainshtein-Zakharov (KSVZ) axion model~\cite{GrillidiCortona:2015jxo}. 

\begin{figure} [t!]
\centering
    \includegraphics[width=1\columnwidth]{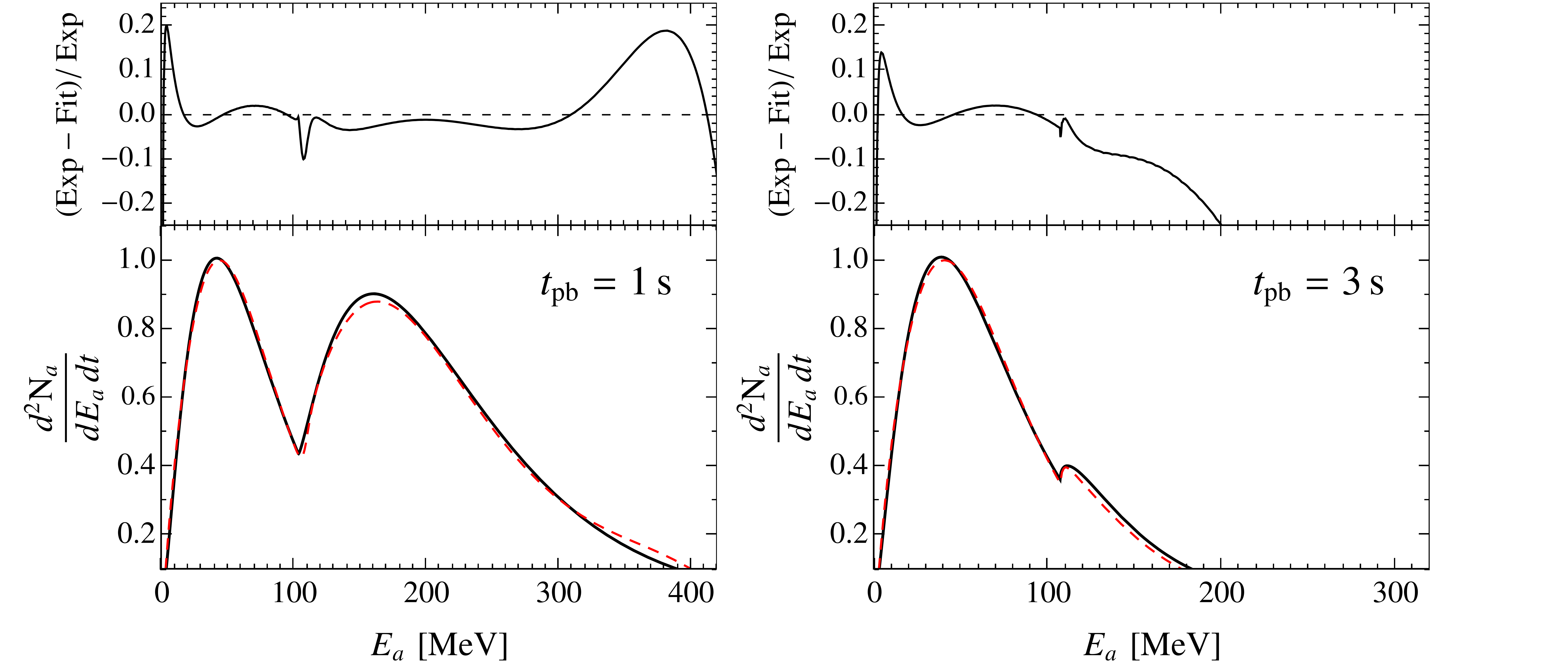}
    \caption{ALP spectrum produced by nuclear processes together with their analytical fits at $t_{\mathrm{pb}}=1\,\s$ (left panels) and $t_{\mathrm{pb}}=3\,\s$ (right panels). All the curves are normalized at their peak value. In particular, the dashed red line refers to the numerical integration over the SN profile, while the solid black line depicts the line shape obtained through the fitting expressions in Eq.~\eqref{eq:BremFit} and Eq.~\eqref{eq:PionFit}. The upper panels show the values of the residuals in the energy range considered.}
	\label{fig:Fits}
\end{figure}

Since $NN$ bremsstrahlung is a quasi-thermal process (see Ref.~\cite{Lella:2023bfb}), it is possible to employ a Gamma-distribution function~\cite{Payez:2014xsa} to interpolate the behavior of the volume-integrated emission rate for our benchmark SN model 
\begin{equation}
    \left(\frac{d^2N_{a}}{dE_a\,dt}\right)_{NN}=A_{NN}\,\left(\frac{g_{ap}}{5\times 10^{-10}}\right)^2\,\left(\frac{E_a}{E_{NN}^0}\right)^{\beta_{NN}} \exp\left[{-(\beta_{NN}+1)\,\frac{E_a}{E^0_{NN}}}\right]\,,
\label{eq:BremFit}
\end{equation}
where $E_a$ is the red-shifted ALP energy observed at large distances from the SN core, and $A_{NN}$, $E^0_{NN}$ and $\beta_{NN}$ are free fitting parameters. The expression in Eq.~\eqref{eq:BremFit} describes a quasi-thermal spectrum with average energy $E_{NN}^0$ and spectral shape parameter $\beta_{NN}$ ($\beta_{NN} = 2$ describes a perfectly thermal spectrum of ultra-relativistic particles).\\
On the other hand, pion conversions are non-thermal processes. Indeed, the energy carried by the incoming pion is completely transferred to the emitted ALP since nucleons in the initial and final states are non-relativistic~\cite{Lella:2022uwi}.  Nevertheless, assuming pions to be in thermal equilibrium in the core, the pion total energy can be estimated as $E_\pi\sim m_\pi+3T\sim E_a$. The local emission rate for this process shows a lower cutoff in coincidence with the pion mass $\,m_\pi$. However, energies measured at large distance from the emission site suffer from red-shift effects induced by the strong gravitational potential around the SN core. Then, the observed lower cutoff energy is given by $\omega_c\simeq\alpha\,m_\pi$, where $\alpha$ is the lapse factor taking into account gravitational effects. 
Since $\alpha$ is in general not constant in the SN volume and at different post-bounce instants, we will leave $\omega_c$ as a free fitting parameter. Thus, we effectively model the ALP spectrum for pion conversions through a Gamma-distribution function  shifted by an amount of energy equal to $\omega_c$.
Consequently, the fitting formula for the spectrum associated with pionic Compton-like processes can be written as 
\begin{equation}
    \left(\frac{d^2N_{a}}{dE_a\,dt}\right)_{\pi N}=A_{\pi N}\,\left(\frac{g_{ap}}{5\times 10^{-10}}\right)^2\,\left(\frac{E_a-\omega_c}{E^0_{\pi N}}\right)^{\beta_{\pi N}} \exp\left[{-(\beta_{\pi N}+1)\,\frac{E_a-\omega_c}{E^0_{\pi N}}}\right]\,,
\label{eq:PionFit}
\end{equation}
with $A_{\pi N}$, $E^0_{\pi N}$, $\beta_{\pi N}$ and $\omega_c$ free fitting parameters.
In our analysis, we have considered SN profiles referred to the first 8 seconds after the core bounce $t_{\mathrm{pb}}\in[1,8]\,\s$ at time steps of $1\,\s$. The fitting parameters in this time interval for our benchmark SN model SFHo-s18.8 and for the other SN models SFHo-s20 and LS220-s20 are reported in Appendix~\ref{app:Parameters}. Remarkably, the time behavior of the parameters $\{E^0_{NN}, \beta_{NN}, A_{NN} \}$ and $\{E^0_{\pi N}, \beta_{\pi N}, A_{\pi N} \}$ can be fitted through analytic expressions of the form ${\sim t^{-\alpha}\,e^{-d\,t}}$~[see Appendix~\ref{app:Parameters} for details]. Fig.~\ref{fig:Parameters} displays the trend of the fitting parameters for our benchmark SN model SFHo-s18.8 with the corresponding 2-$\sigma$ bands. The monotonic decrease in the $E^0_{NN}$ and $E^0_{\pi N}$  parameters, related to the average energy associated to the considered process, follows the drop in temperature characterizing the SN cooling phase. We also point out that for $t_{\rm pb}\lesssim2\,\s$ we find $\beta_{NN}\simeq1.5$, confirming that ALPs produced via $NN$ bremsstrahlung follow a \emph{quasi}-thermal distribution.\\
As the SN temperature drops, pions in the core are subject to condensation effects. In particular,  Bose-Einstein condensation starts to become relevant in the whole PNS core at $t_{\rm pb}\gtrsim 3\,\s$. By setting pion conversion rates to zero in the regions affected by bosonic condensation, ALP production from pionic processes can occur just from the outer layers of the core where the temperature is lower, leading to a strong suppression of this production channel. This effect can clearly be observed from the behavior of the pion conversion normalization constant $A_{\pi N}$, shown in the lower-right panel of Fig.~\ref{fig:Parameters}.
Fig.~\ref{fig:Fits} displays two examples of fits performed at two different instants. The goodness of the employed fitting formulae is highlighted by the values of the residuals, shown in the upper panels of Fig.~\ref{fig:Fits}. In particular, the discrepancy between the numerical fluxes and the fitted ones is never larger than 20\%. Moreover, it is possible to appreciate that the drop in temperature and the rise of condensation effects lead to significant suppression of the pion conversion contribution at $t_{\mathrm{pb}}\gtrsim3\,\s$.

\subsection{One-zone model}
\label{subsec:onezone}

\begin{figure} [t!]
\centering
    \includegraphics[scale=0.7]{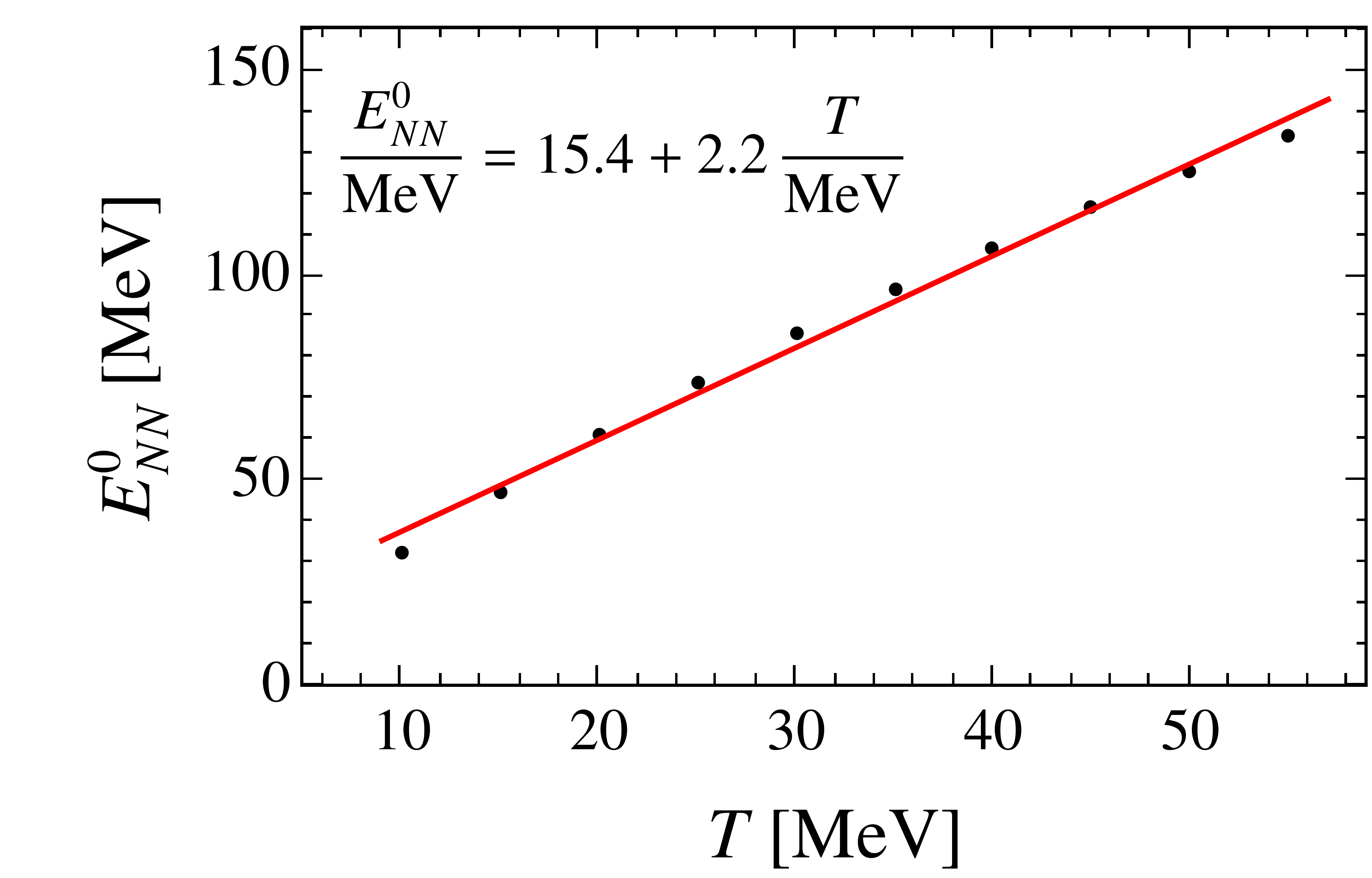}
    \caption{Values of $E^0_{NN}$ for the fitting spectra obtained from our schematic one-zone SN model at different temperatures. The red solid line depicts the linear fit in the temperature range ${T\in[10,55]~\MeV}$.}
	\label{fig:E0vsT}
\end{figure}

The shape of the ALP emission spectra is strictly dependent on the temperature of the inner regions of the PNS at $R\lesssim 10\,\km$, where nuclear processes take place. This is due to the fact that only in these regions nuclear densities and temperatures are high enough to induce efficient ALP production.  On the other hand, neutrinos contributing to the SN cooling are emitted from a last-scattering surface,  usually called \textit{neutrino-sphere}~\cite{Raffelt:1996wa}. In particular, during the cooling phase at $t_{\rm pb}>1\,\s$ the neutrino-sphere is typically located at $R_\nu\sim 30\km$~\cite{Janka:2006fh}.
Therefore, if the analysis of the SN neutrino spectra can provide information about the temperature of the outer regions of the PNS, ALPs could be exploited as thermometers of the inner core. 
In order to establish how the fitting parameters depend on the temperature of the core, we have fitted through Eq.~\eqref{eq:BremFit} and Eq.~\eqref{eq:PionFit} the spectra obtained from an exemplary SN toy-model, in which the temperature is assumed to be constant for $R\leq13\,\km$ and null elsewhere. In the same way, the mass density and the electron fraction are set to $\rho=3\times10^{14}\,\g\,\cm^{-3}$ and $Y_e=0.3$ in this region, while they are assumed to be null in the outer regions. Notice that the value of the radius of this \textit{one-zone} model is chosen to have a PNS of mass $M_{\text{NS}}\simeq1.4\,M_\odot$, which is a typical value for the compact object at the center of an exploding SN~\cite{Raffelt:1996wa}. All the other SN parameters are obtained as outputs of the EoS employed to describe the nuclear medium. In particular, we use as a benchmark the SFHo EoS. Then, by varying the temperature of the one-zone SN model in the range ${T\in[10,55]\,\MeV}$, we have determined the fitting parameters for each temperature. In particular, in analogy to the neutrino \emph{quasi}-thermal spectra, we expect the fitting parameter $E^0_{NN}$ to show a linear dependency on the average temperature in the production region~\cite{GalloRosso:2017mdz,Lujan-Peschard:2014lta}. As shown in Fig.~\ref{fig:E0vsT}, in the temperature range of interest, the correlation between $E^0_{NN}$ and $T$ can be expressed as
\begin{equation}
    \frac{E^0_{NN}}{\MeV} = 15.4 + 2.2 \frac{T}{\,\MeV}\,.
    \label{eq:E0B}
\end{equation}\newline
As we will see in the following Section, ALPs convert in the magnetc field of the Galaxy leading to an observable gamma-ray burst whose spectral and temporal behaviors do encode information also on $E^{0}_{{NN}}$. The reconstruction of $E^{0}_{{NN}}$  from gamma-ray data can therefore provide an immediate estimation of the PNS temperature.
Nevertheless, the energies observed at large distance from the production point suffer from the gravitational red-shift, encoded in the lapse factor $0<\alpha<1$. Therefore, in this procedure the observed energy has to be multiplied by a factor $1/\alpha$ to recover the local energy. Thus, by reverting Eq.~\eqref{eq:E0B}, the temperature can be estimated as
\begin{equation}
    \frac{T}{\MeV}\simeq-6.93+\frac{0.45}{\alpha}\,\left(\frac{E^{0}_{NN}}{\,\MeV}\right)_{\rm obs} \,\ .
\label{eq:TRecForm}
\end{equation}\newline
We highlight that the lapse factor $\alpha$ is in general not constant in the inner PNS, where gravitational effects are relevant, and a complete general relativistic treatment is necessary. However, the lapse function does not vary significantly in regions where ALP production is efficient at $R\sim5-10\,\km$, thus for this estimation it is possible to assume the average value assumed by $\alpha$ in these regions. In particular, we will use the value $\alpha\approx0.78$ for the SFHo-s18.8 model and $\alpha\approx0.67$ for the SFHo-s20 and LS220-s20 models. The difference in the values assumed is due to the different PNS masses characterizing these models. Since the latter models lead to higher PNS masses~(see Sec.~\ref{sec:SNmodels}), gravitational effects are more relevant and, thus, $\alpha$ is smaller.
We highlight that, since the value of $\alpha$ depends on the PNS mass, the determination of this parameter represents an additional source of uncertainty for our analysis. However, a rough estimation of the mass of the PNS associated with a future Galactic SN can be recovered by looking at the neutrino signal, as explained in Ref~\cite{GalloRosso:2017hbp}. 

\section{ALP conversions in Galactic magnetic fields}
\label{sec:Conversions}

Once produced in the interior of the PNS, ultra-light ALPs (with $m_a\lesssim 0.1$~neV) may convert into photons
thanks to $g_{a\gamma}$, while propagating into Galactic magnetic fields.
Thus, given the ALP-photon conversion probability $P_{a\gamma}(E)$, the induced photon spectrum can be obtained as
\begin{equation}
    \frac{d^2\phi_\gamma}{dE\,dt}=\frac{1}{4\pi d^2} \frac{d^2N_{a}}{dE\,dt}P_{a \gamma}(E)\,,
    \label{eq:spec1}
\end{equation}
where we adopt $d=10\,\kpc$ as fiducial distance for a future Galactic SN event. \\
In principle, Galactic magnetic fields show a complex structure, so that the ALP propagation in the Milky Way requires in principle a truly 3-dimensional treatment (see, e.g.,~\cite{DeAngelis:2011id}). 
In the following we will consider the case of energy independent conversion probabilities, which is recovered for ALP masses~\cite{Calore:2023srn}
\begin{equation} 
   m_a \ll  0.36\,{\rm neV}\left(\frac{E}{100 \,\ {\rm MeV}} \right)^{1/2}\left(\frac{L}{10 \,\ {\rm kpc}} \right)^{-1/2}\,,
    \label{eq:mcrit}
\end{equation}
In this case the ALP conversion probability can be well approximated by  (see Ref.~\cite{Calore:2023srn}, e.g., for further discussion)
\begin{equation}
P_{a\gamma} \simeq (\Delta_{a \gamma}L)^2 \,\,, 
\label{eq:enindep}
\end{equation}
where 
\begin{equation}
        \Delta_{a\gamma}\simeq \, 1.5\times10^{-6} \left(\frac{g_{a\gamma}}{10^{-15}\,\GeV^{-1}} \right)
        \left(\frac{B_T}{10^{-6}\,\rm G}\right) \kpc^{-1}\,.
    \label{eq:Delta0}
\end{equation}
where $B_T$ is the average value of the transverse component of the magnetic field along the line of sight. In our analysis we assume a SN exploding at $10~\kpc$ from Earth in the direction of $(\ell, b) = (199.79^{\circ}, −8.96^{\circ})$, the same as in Ref.~\cite{Calore:2023srn}. This direction in the sky coincides to the location of Betelgeuse, a Red Supergiant among the most promising Galactic SN candidates~\cite{vanLeeuwen:2007tv}. Following the discussion in Ref.~\cite{Calore:2023srn}, we employ the Jansonn and Farrar model~\cite{Jansson:2012pc} with the updated parameters of Ref.~\cite{Planck:2016gdp} (JFnew) as benchmark for the Galactic magnetic field. For this model, $B_{T}=0.58~{\rm \mu G}$ and we can parameterize the conversion probability as   
\begin{equation}
    P_{a\gamma}=10^{-10}\times\left(\frac{g_{a\gamma}}{10^{-15}\GeV^{-1}}\right)^2\,P_0\,,
\label{eq:P0}
\end{equation}
where $P_0=0.785$. Since we are assuming an energy-independent conversion probability, the behavior of the photon flux follows the shape of the ALP emission spectrum. In this case, a different model for the Galactic magnetic field would affect only the value of $P_0$ in Eq.~\eqref{eq:P0} (e.g., for the Pshirkov model~\cite{Pshirkov:2011um} $B_T=0.24~\mu{\rm G}$ and $P_0=0.128$~\cite{Calore:2023srn}). This would imply a different normalization for the photon flux arriving at Earth, without affecting its shape and its average energy. Therefore, in our analysis we take into account only the JFnew model.\\
The value of the conversion probability could depend also on the location of the SN event in the sky. In this regard, we refer the reader to Fig.~7 of Ref.~\cite{Calore:2021hhn}, which describes how the photon flux induced by ALP conversion varies by taking into account different directions. This quantity can be considered barely constant for sources along the Galactic disk and outside of the Galactic center, where many Galactic red supergiants are located~\cite{Healy:2023ovi}. Thus, the reference direction assumed in this work provides a reliable estimation of the photon flux expected for such Galactic SN candidates.

The photon flux observed by the \emph{Fermi}-LAT experiment is additionally altered by instrumental effects. In particular, it suffers from smearing effects due to the finite energy resolution of the detector. 
Thus, the observed gamma-ray flux reads:
    \begin{equation}
    \frac{d^2\phi_{\gamma,\rm obs}}{dE_\gamma \,dt} = \int_{-\infty}^{+\infty} \eta(E,E_\gamma)\,\frac{d^2\phi_\gamma}{dE\, dt}(E)\,dE\;,
    \label{eq:spec2}
\end{equation}
where, we approximate the detector energy resolution as
\begin{equation}
    \eta(E,E_\gamma)=\frac{e^{-(E-E_\gamma)/2\,\sigma^2}}{\sqrt{2\pi\sigma^2}}\,
    \label{eq:Resolution}
\end{equation}
and $\sigma=0.2\,E_\gamma$~\cite{Calore:2023srn}. 
We stress that the previous two equations are meant to convey an intuitive idea of how instrumental effects alter the spectrum of observed photons. 
However, our analysis will be based on full simulations of \fermi-LAT observation, as described in Sec.~\ref{sec:FermiLAT-analysis}.

\begin{figure} [t!]
\centering
    \includegraphics[width=1\columnwidth]{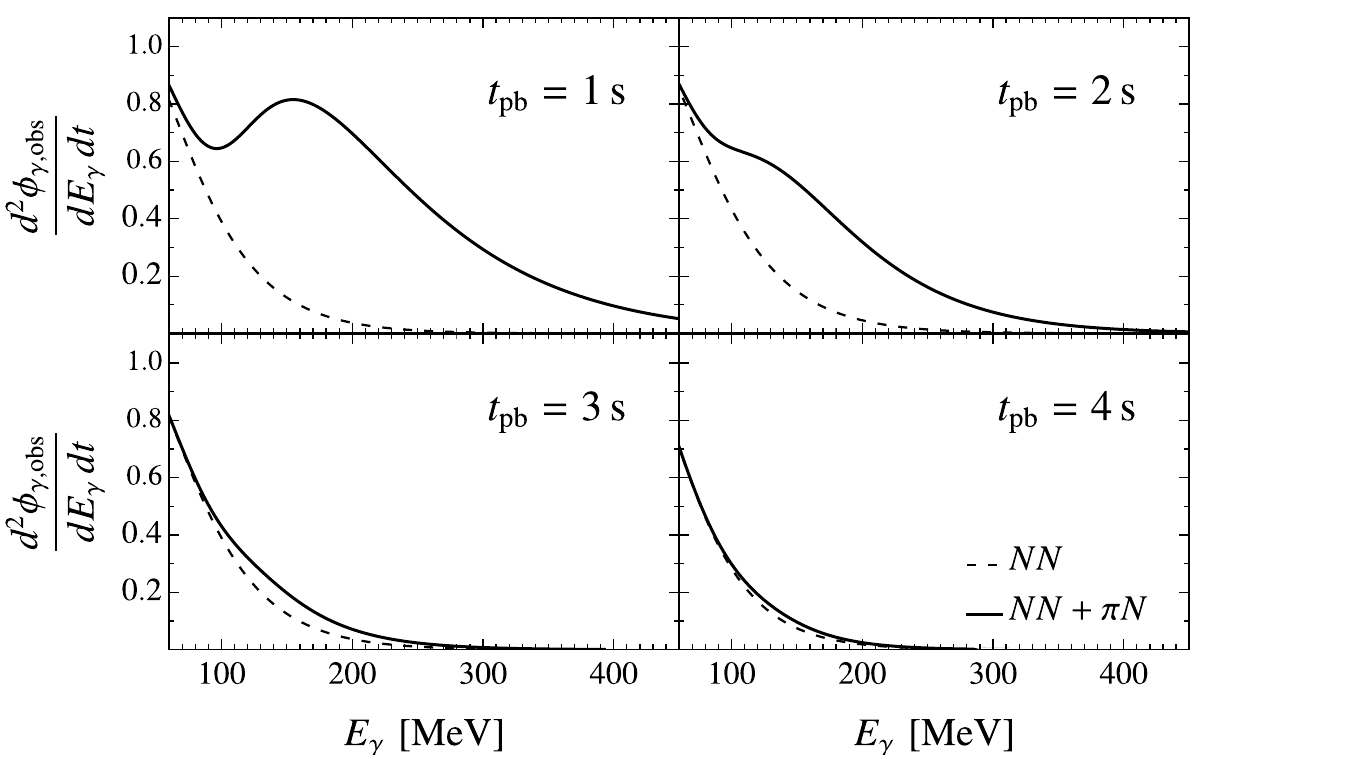}
    \caption{Normalized gamma-ray spectra smeared by the \fermi-LAT energy resolution in Eq.~\eqref{eq:Resolution}. ALP emission is set on our benchmark SN model SFHo-s18.8. In particular, the four panels refer to different snapshots of the signal at the first 4 seconds after the core bounce. Moreover, the solid line depicts the spectrum in presence of pions ($NN+\pi N$), while the dashed line is the spectrum induced by bremsstrahlung alone ($NN$).}
	\label{fig:SmearedSpectra}
\end{figure}

Fig.~\ref{fig:SmearedSpectra} displays four snapshots of the predicted photon signal observed by the \fermi-LAT experiment at four different times after the core bounce. Notice that the photon flux is normalized at its maximum value, in order to make it independent on the SN distance and on the ALP-nucleon coupling $g_{aN}$ and the ALP-photon coupling $g_{a\gamma}$. We consider only the energy range $E\gtrsim60\,\MeV$, in which the \fermi-LAT effective area is large enough to permit the detection of the signal~\cite{Calore:2023srn}.
Indeed, as we will see in the next Section, in a full simulation of the observed events through the \fermi-LAT instrument response convolution, the effective area plays a crucial role in shaping (and altering) the peak of the $NN$ component at low energies.
We highlight that for $t_{\rm pb}\lesssim2\,\s$, when pion conversion is still efficient, smearing effects do not cancel an eventual bump associated to pionic processes. 
Thus, the \fermi-LAT energy resolution is good enough to discriminate if pion conversion is a viable ALP production channel in the SN core.

\section{\fermi-LAT gamma-ray signal analysis}
\label{sec:FermiLAT-analysis}

We want to go beyond the considerations above that invoked effective treatments of the \fermi-LAT response functions. Our goal is to simulate the ALP-induced gamma-ray spectra and reconstruct their parameters in a \fermi-LAT data analysis chain following the procedure explained in Ref.~\cite{Calore:2023srn}. The framework developed therein allows us to generate the prompt gamma-ray burst due to ALP production in a core-collapse event. We slightly adapt the methodology to our needs for the study of $NN$ and $\pi N$ ALP interactions in a SN core. In what follows, we list the main ideas and the modifications of the analysis approach fully described in Ref.~\cite{Calore:2023srn}. We point the interested reader to the latter publication for the omitted details. 

\paragraph*{Analysis setup.} The target SN event takes place at 10 kpc distance from Earth in the direction of $(\ell, b) = (199.79^{\circ}, −8.96^{\circ})$. The expected ALP-induced gamma-ray signal, either originating in $NN$ bremsstrahlung or Compton-like $\pi N$ processes, is simulated for 8 seconds with the Fermi Science Tools.\footnote{\url{https://fermi.gsfc.nasa.gov/ssc/data/analysis/software/}} Due to the transient nature of the event, we perform all simulations for the \texttt{P8R3\_TRANSIENT020\_V3} event class and \texttt{FRONT+BACK} type events while setting the explosion time to $T_{\mathrm{ON}} = 510,160, 000$ MET, a representative ON-time of the LAT. The simulation is performed for energies between 60 to 600 MeV with an additional cut on the zenith angle $<80^{\circ}$ of the reconstructed events.

\paragraph*{Gamma-ray model preparation.} Both the instrument response functions of \fermi~LAT and all quality cuts on the reconstructed events have a significant impact on the spectral profile of the gamma-ray burst. It is crucial to know how the theoretically-expected signature would look like in reality through the ``lens'' of an actual experiment. To enable a determination of the SN core temperature, it is also essential to study the time-dependent gamma-ray spectrum of the transient event. This requires a detailed simulation of the theoretical model introduced in Sec.~\ref{sec:ALP production}. For this purpose, we process the time-dependent parametric fit formula for $NN$ bremsstrahlung events in Eq.~(\ref{eq:BremFit}) with the \fermi~Science Tools using a $35\times35$ grid for the parameters $\beta_{NN}\in\left[0.5, 2.5\right]$ and $E_{NN}^0\in\left[50, 200\right]$ MeV. We set the normalization of the spectrum to $A_{NN} = 100\times\Tilde{A}_{NN}$ 
and generate 30 realizations of these grids for an observation time of $1\,\mathrm{s}$ to eventually derive an average \fermi-LAT model $\bm{\mu}_{NN}$. The average spectrum is split into 20 logarithmically spaced energy bins between 60 and 600 MeV. This leaves us with a model incorporating the data selection cuts and, in particular, the energy dispersion of the LAT. Since the ALP fit formula is valid in good approximation for each time step after the core collapse, we can fit any observation at any time with our model and reconstruct the parameters. As the model reflects the average expectations, we simply need to linearly re-scale the normalization to fit any observed spectrum. 

In full analogy, we prepare a \fermi-LAT model $\bm{\mu}_{\pi N}(t)$ of pionic processes based on Eq.~(\ref{eq:PionFit}). This depends on a fourth parameter $\omega_c$ and therefore needs to be prepared per time interval. To simplify the simulation work, we fix $\omega_c$ and its time-dependence to the respective prediction from the numerical SN simulation with model SFHo-s18.8. Consequently, we simulate the formula on a grid of $E^0_{\pi N}\in\left[100, 400\right]$ MeV and $\beta_{\pi N}\in\left[0.2, 4\right]$ for time intervals of $1\,\mathrm{s}$ with $A_{\pi N} = 100\times\Tilde{A}_{\pi N}$. 

\begin{figure} [t!]
\centering
    \includegraphics[width=1\columnwidth]{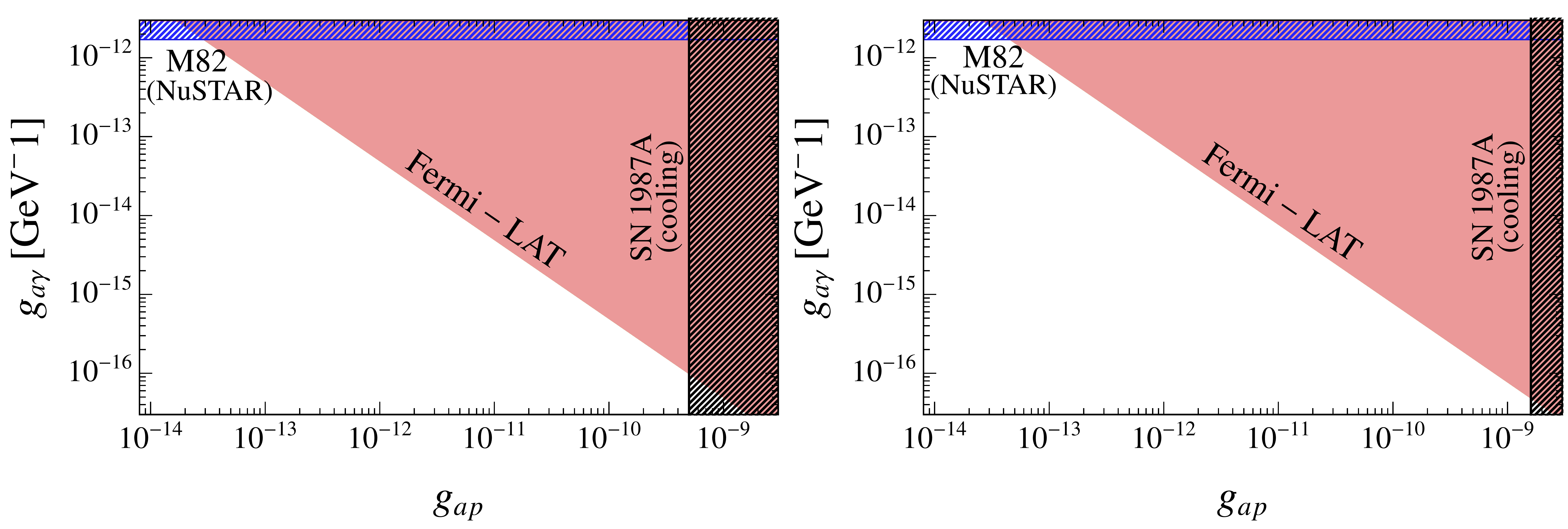}
    \caption{\fermi-LAT sensitivity (red) in the plane $g_{a\gamma}$ vs $g_{ap}$ for the Galactic SN event under consideration. The physics case considering the presence of pions in the SN core is shown in the left panel, while the right panel refers to the scenario in absence of pions. For each scenario, we report in hatched black limits on $g_{ap}$ from SN 1987A cooling~\cite{Lella:2023bfb} and in hatched blue constraints from NuSTAR observations of the M82 galaxy~\cite{Ning:2024eky}.}
	\label{fig:Sensitivity}
\end{figure}

\paragraph*{Parameter reconstruction.} We use a Poisson likelihood function per energy bin 
\begin{equation}
\label{eq:logP}
\mathcal{L\!}\left(\left.\bm{\mu}\right|\bm{n}\right) = \prod_{i=1}^{N_E}  \frac{\mu_i^{n_{i}}}{\left(n_{i}\right)!}e^{-\mu_i}
\end{equation}
as the statistical model for our parameter inference with \texttt{MultiNest} \cite{Feroz:2008xx}, where $N_E$ is the number of energy bins. We use 1000 live points and an evidence tolerance of 0.2 following the recommendations of the developers for parameter estimation tasks. The target observation $\bm{n}$ is produced with the exact time-dependence of the parameters for each contribution, respectively, given in Appendix~\ref{app:Parameters}. To simulate the target data, we choose a combination of ALP couplings in the reach of \fermi-LAT and compatible with currently existing bounds from astrophysical observations, i.e.~${g_{a\gamma} = 3\times10^{-15}}~\GeV^{-1}$ and $g_{ap} = 5\times10^{-10}$. Following the procedure described in Sec.~III.B of Ref.~\cite{Calore:2023srn}, we compute the \fermi-LAT sensitivity for the Galactic SN case under examination. The result is shown as a red region in Fig.~\ref{fig:Sensitivity} in the plane $g_{a\gamma}$ vs $g_{ap}$ considering an ALP mass $m_a=10^{-10}\,\eV$. We also report the leading astrophysical constraints on $g_{ap}$~\cite{Lella:2023bfb} (hatched black region) and $g_{a\gamma}$~\cite{Ning:2024eky} (hatched blue). In particular, the LAT sensitivity scales as $\propto g_{ap}^2 \times g_{a\gamma}^2$. For $g_{ap}=5\times10^{-10}$ \fermi-LAT is sensitive to ALP-photon couplings $g_{a\gamma}\gtrsim0.97\times10^{-16}\,\GeV^{-1}$ in presence of pions in the SN core (left panel) and to $g_{a\gamma}\gtrsim1.52\times10^{-16}\,\GeV^{-1}$ in absence of pions in the SN core~(right panel). We remark that the value of the mass does not affect the reported sensitivity as long as $m_a\ll10^{-9}\,\eV$~(see Eq.~\eqref{eq:mcrit}). Analogously, bounds on $g_{ap}$ are not sensitive to the small value of the mass, while constraints on $g_{a\gamma}$ slightly change with the mass in this range.

\section{Results}
\label{sec:Results}
\subsection{Presence of pions in the SN core}
\begin{figure} [t!]
\centering
    \includegraphics[width=0.49\columnwidth]{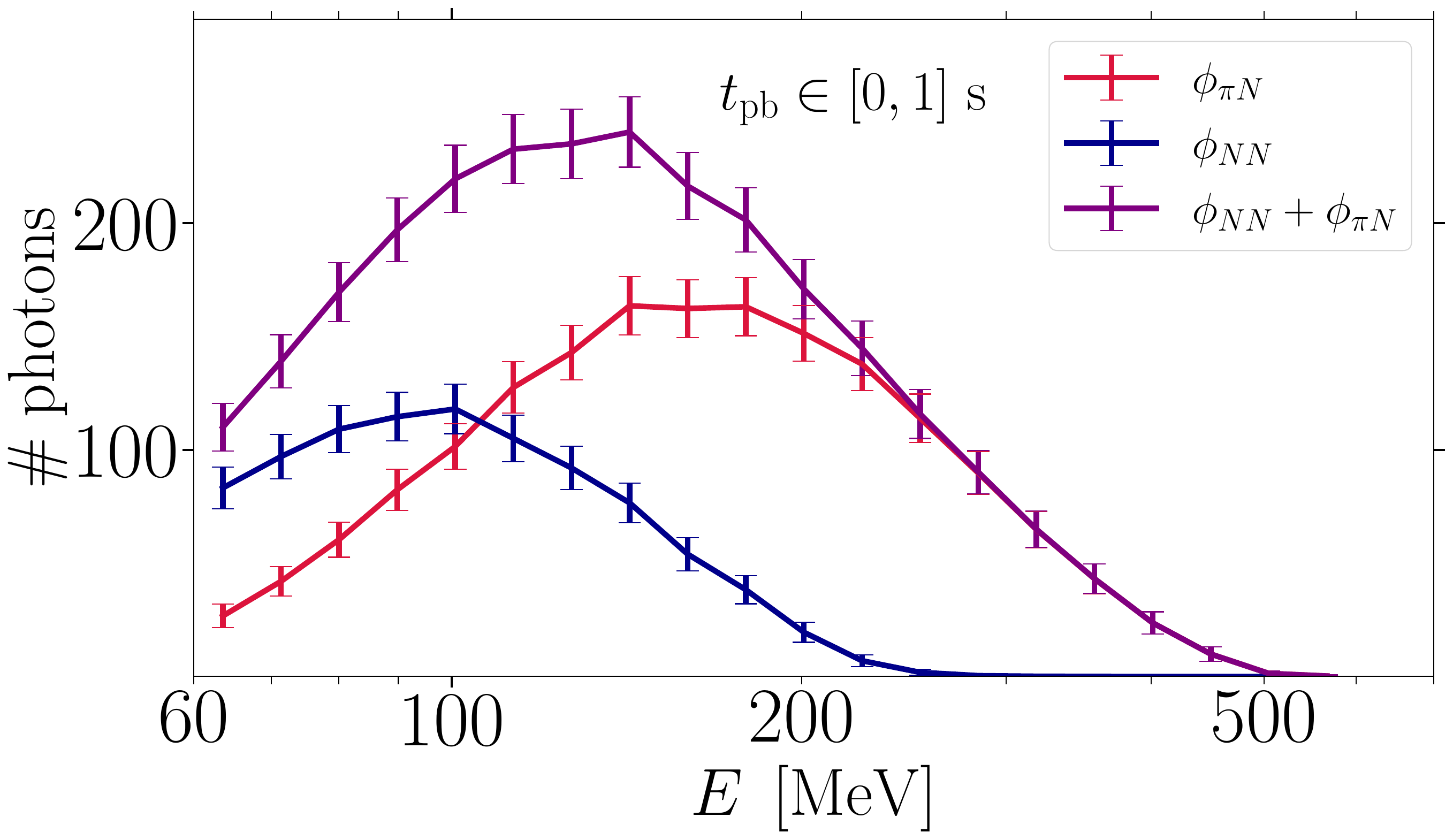}\hfill    \includegraphics[width=0.49\columnwidth]{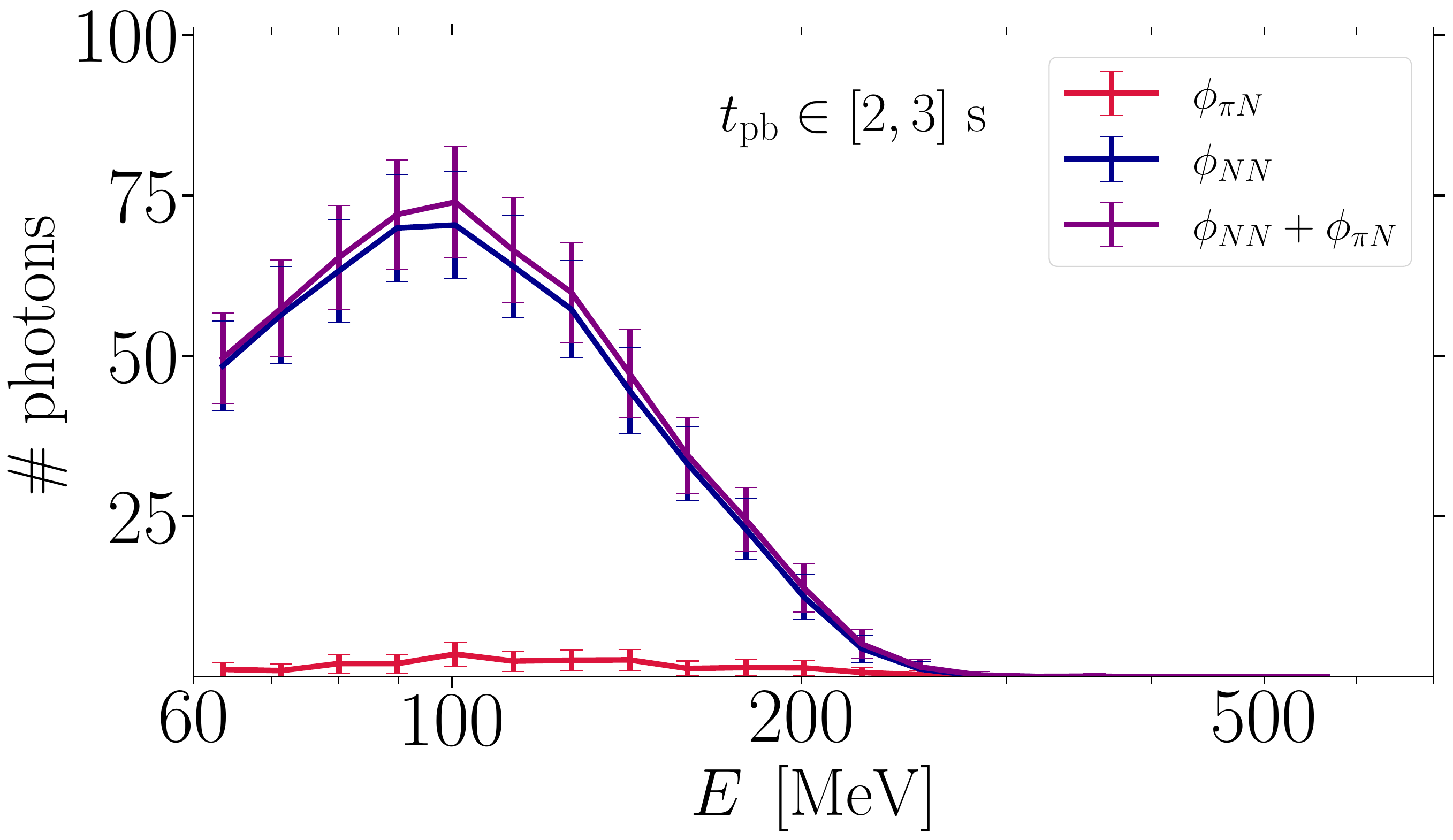}
    \caption{\textit{Fermi}-LAT spectra of the ALP-induced gamma-ray burst from a core-collapse SN in the time slices $t_{\rm pb}\in\left[0, 1\right]$ s (\textit{left}) and $t_{\rm pb}\in\left[2, 3\right]$ s (\textit{right}) after the bounce. We fixed ${g_{a\gamma}=3\times10^{-15}~\GeV^{-1}}$ and $g_{ap} = 5\times10^{-10}$ to compute the average expectation for the number of gamma-ray photons associated with the bremsstrahlung (blue) and pion conversion (red) components. The sum of both contributions is shown in purple. The simulation conditions follow the details provided in Sec.~\ref{sec:FermiLAT-analysis}.}
	\label{fig:ALPspectrum_in_FERMI}
\end{figure}

As explained in Sec.~\ref{sec:intro}, the observation of an enhancement in ALP emission in the range of energies characterizing pionic Compton-like processes would represent striking evidence supporting the scenario discussed in Ref.~\cite{Fore:2019wib}. \newline
Figure~\ref{fig:ALPspectrum_in_FERMI} depicts the simulated behavior of the signal in the \fermi-LAT experiment related to an ALP burst from a  future Galactic SN  at distance $d=10\,\kpc$.  The snapshot considered in the left panel refers to the first second time-bin after the core-bounce. The contribution due to $NN$ bremsstrahlung is peaked at energies $E\sim90\,\MeV$, just above the \fermi-LAT energy threshold. Indeed, the \fermi-LAT experiment is characterized by a dramatic reduction in the effective area for the detection in the energy range $E\lesssim80\,\MeV$. This results in a significant suppression of the bremsstrahlung emission spectrum with respect to the spectral shape depicted in Fig.~\ref{fig:Fits}. On the other hand, the pion conversion peak is well resolved in the energy range in which it is efficient $E\sim150-200\,\MeV$. Thus, the presence of a relevant $\pi N$ component can be pointed out by observing a significant number of detected photon events in the range of energies $E\gtrsim200\,\MeV$, where the contribution from bremsstrahlung is expected to be null. Therefore, this feature can provide striking evidence for the presence of a non-negligible pion population inside the PNS. In the right panel of Fig.~\ref{fig:ALPspectrum_in_FERMI}, we also show the evolution of the burst's spectrum in the time interval $t_{\rm pb}\in[2,\,3]$ s. In this phase of the gamma-ray burst, the additional contribution due to pion conversion ceases to be significant and the emission is dominated by the bremsstrahlung component. Hence, already after a few seconds, employing only the bremsstrahlung contribution is an excellent description of the physical process.
Since the bremsstrahlung contribution is suppressed at energies below 100 MeV by the LAT's effective area, we do not see the double-peak shape of the ALP spectrum shown in Fig.~\ref{fig:AxionProduction} translated into gamma rays, and yet this ``observed'' spectrum carries information about both production mechanisms.

\medskip 

Since the presence of gamma rays above 200 MeV in the first second of the burst is a good indicator of the presence of pionic processes in the SN core, we can quantify the preferred physical nature of the emission given the two fit formulae. Our Bayesian framework allows for model comparison based on the so-called \textit{Bayes factor}. It is defined by virtue of the Bayesian evidence $\ln{\mathcal{H}_X}$ of hypothesis/model $X$ and the corresponding evidence $\ln{\mathcal{H}_Y}$ of an alternative hypothesis/model $Y$. The Bayes factor then reads $\mathcal{B}_{XY} = \exp{(\ln{\mathcal{H}_X} - \ln{\mathcal{H}_Y})}$. A positive Bayes factor means that model $X$ is preferred over model $Y$ to a certain degree by the dataset under scrutiny. In contrast to the frequentist point of view, there is no absolute significance of the preference associated with a certain Bayes factor; it is rather an empirical classification. We refer to the degree of evidence from Table 1 of Ref.~\cite{Trotta:2008qt} based on the logarithmic Bayes factor to interpret the numerical results.

Applied to the combined spectrum shown in Fig.~\ref{fig:ALPspectrum_in_FERMI}, a fit with only the bremsstrahlung component yields a (log-)Bayesian evidence of $\ln{\mathcal{H}_{NN}} = 5529$ while a fit with a model accounting for both components yields a value of $\ln{\mathcal{H}_{NN+\pi N}} = 5535$. Thus, we obtain a Bayes factor of $\ln{\mathcal{B}} = -6$ implying that there is a strong evidence for the combined model being a better fit to the data.

Therefore, we have demonstrated that, in the presence of $\pi N$ and for sufficient $g_{aN}$ couplings, a Bayesian analysis can
detect the evidence for this additional contribution on top of the $NN$ one. In particular, the pion conversion contribution can be identified by the sizable abundance of photon events above 200 MeV within the first second(s) of the burst. 

For the sake of completeness, we point out that a detectable gamma-ray burst at energies $E_a\gtrsim200\,\MeV$ might be produced also in case of heavy ALPs with masses $m_a\gtrsim100\,\MeV$ thermally produced in the SN core and decaying in photon pairs along their path from the exploding SN to the Earth~\cite{Jaeckel:2017tud,Hoof:2022xbe,Muller:2023pip, Muller:2023vjm}. Nevertheless, the two scenarios could be easily disentangled. In particular, ALPs with masses $m_a\sim200\,\MeV$ decay in photons over time scales of tens of minutes for the values of the couplings considered in this work. In this time window, heavy ALPs propagate much slower than the ultralight pseudoscalars considered in this work. Therefore, the ALP burst in this ``heavy-ALP scenario'' would be characterized by a huge time delay with respect to the observation of the SN neutrino burst. Conversely, in the ``light-ALP scenario'' the detection of SN neutrinos and SN ALPs would occur simultaneously. Finally, we highlight that the heavy-ALP scenario is disfavored with respect to the light-ALP one, since the region of the parameter space $(m_a,g_{a\gamma}, g_{ap}) \sim (200\,\MeV, 10^{-15}\,\GeV^{-1}, 10^{-10})$ is robustly excluded by observations of the diffuse SN ALP background~(see Fig.~4 of Ref.~\cite{Lella:2022uwi}).

\begin{figure} [t!]
\centering
\includegraphics[width=0.95\textwidth]{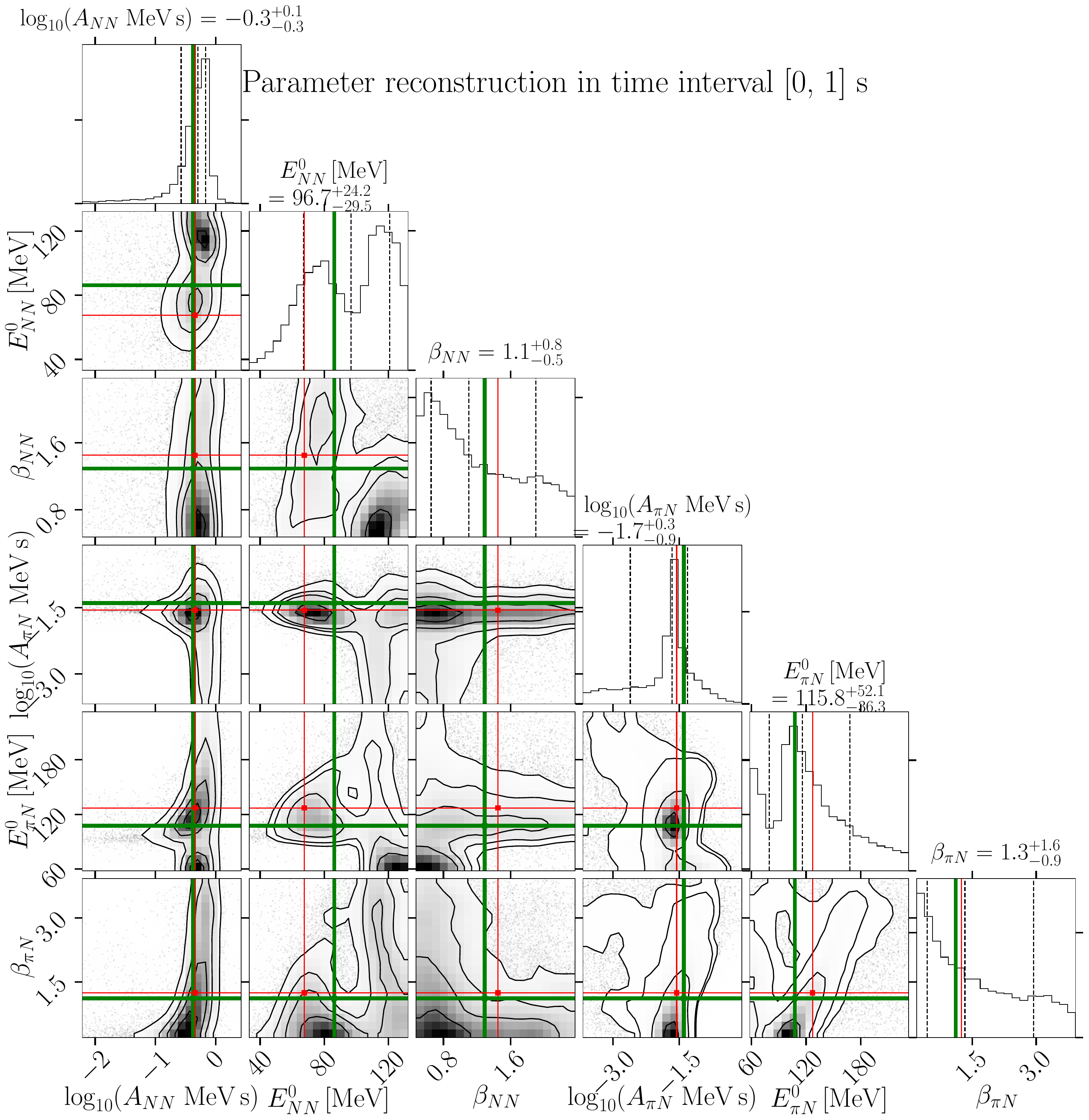}
    \caption{Best-fitting values and posterior distributions of the reconstructed ALP spectral parameters $A_{NN}, E_{NN}^0$ and $\beta_{NN}, A_{\pi N}, E_{\pi N}^0$ and $\beta_{\pi N}$ from the ALP-induced gamma-ray spectrum observed within the first second after the core bounce. The burst receives contributions from bremsstrahlung and pion conversion processes computed employing our benchmark model SFHo-s18.8. We assume a light ALP $m_a \ll 0.1$ neV, $g_{a\gamma} = 3\times10^{-15}$ GeV$^{-1}$ and $g_{ap} = 5 \times 10^{-10}$. We overlay the marginal two-dimensional posterior distributions with the parameter values used to simulate the mock signal in red, while the green values denote the parameter values maximizing Eq.~\eqref{eq:logP}. The marginal one-dimensional posterior distributions for each parameter show the $16\%$, $50\%$ (median) and $84\%$ quantiles as dashed black lines, whose numerical values are also stated in the title of each marginal posterior.}
\label{fig:posteriors_paramReco_Brems+pion_dT_1s}
\end{figure}

\subsection{Reconstruction of ALP spectra fitting parameters}
\label{sec:LAT-ParamReconstruction}
As explained in Sec.~\ref{sec:TemperatureReconstruction}, reconstructing the PNS temperature and its temporal evolution is tied to the possibility to reconstruct the spectral parameters of the gamma-ray burst, and, most notably, $E_{NN}^0$. 
In this Section, we attempt the reconstruction of the parameters of the ALP spectrum from the analysis of \fermi-LAT data. 
In our Bayesian formalism, outlined in Sec.~\ref{sec:FermiLAT-analysis}, we can indeed also infer the full posteriors of all fitted parameters. Our particular interest is focused on the reconstruction of the mean energy $E_{NN}^0$. 

The full model of the gamma-ray burst is characterized by six spectral parameters associated with the bremsstrahlung and pionic contribution to the ALP production. In Fig.~\ref{fig:posteriors_paramReco_Brems+pion_dT_1s} we show a representative example of the parameter inference results for the SFHo-s18.8 model. Here we consider the gamma-ray burst spectrum obtained from the first second after the core bounce. We display the two-dimensional and one-dimensional marginal posterior distributions for all six parameters. The red lines and dots indicate the true, injected values of the ALP signal while the green counterparts denote the \textit{mean} reconstructed values. Despite the evidence for the pion conversion
component we reported in the previous section, large degeneracies remain between the model's parameters. Even worse, the posterior of $E_{NN}^0$ exhibits a bi-modal structure. An explanation for this behavior may be the apparent similarity of both spectral components (the one from bremsstrahlung and the one from pion conversion) after passing the theoretical models through the Fermi Science Tools and applying data selection cuts to the generated photon events (see the left panel of Fig.~\ref{fig:ALPspectrum_in_FERMI}). In fact, $E_{\pi N}^0$ exhibits a similar bi-modal shape indicating that both components can be interchanged. Therefore, it does not seem feasible to achieve a satisfactory reconstruction of $E_{NN}^0$ in the presence of a strong pion conversion component, at least in the first second after the onset of the burst.

\begin{figure} [t!]
\centering
    \includegraphics[width=0.8\textwidth]{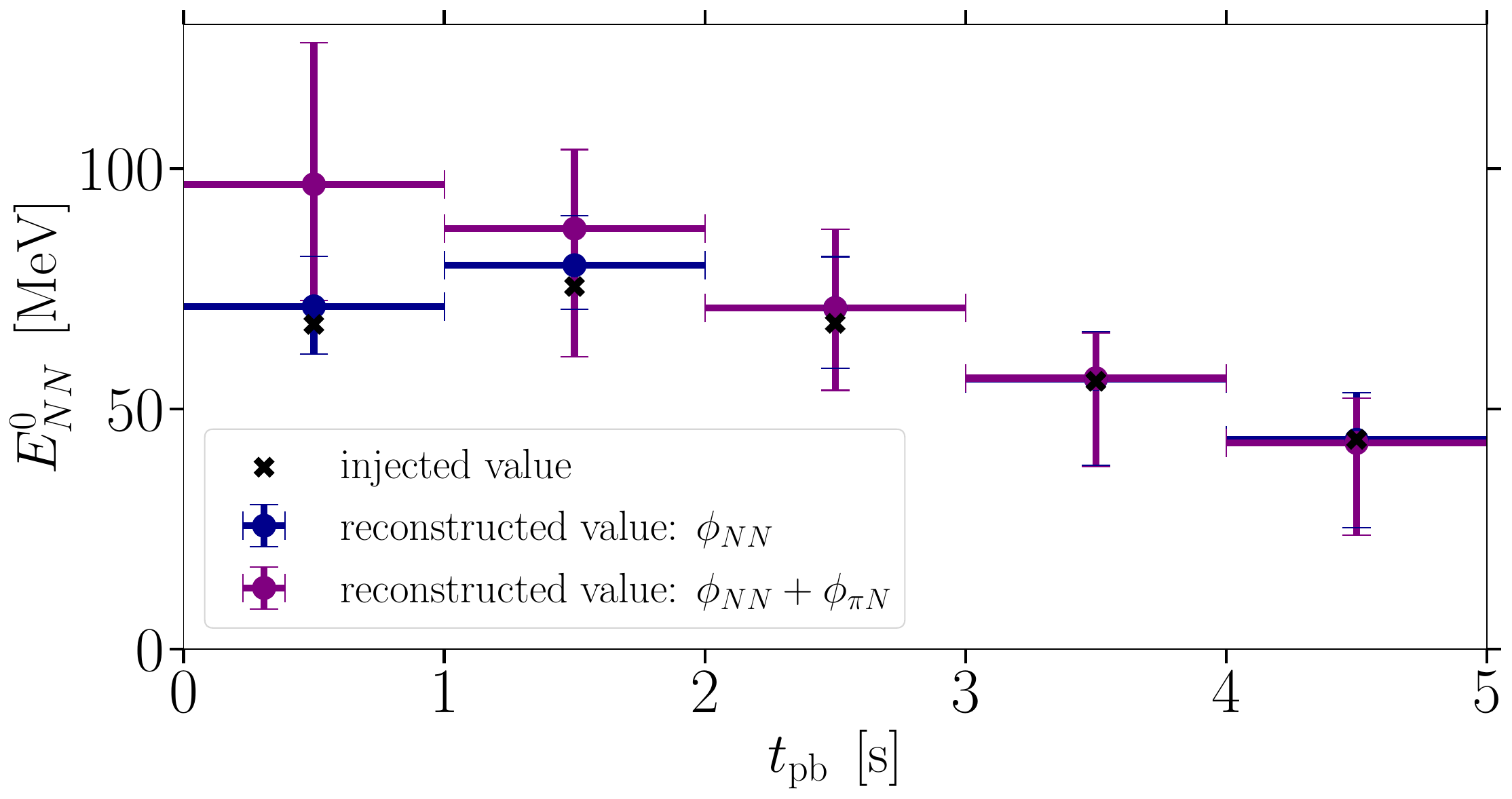}
    \caption{Time evolution of $E_{NN}^0$ as reconstructed from a simulation of the SFHo-s18.8 SN model for up to five seconds after the core bounce. The injected values of the underlying model are marked with a black cross. We distinguish the two cases of a gamma-ray burst comprised of the summed emission of bremsstrahlung and pion conversion processes (purple) as well as only bremsstrahlung. The vertical error bars indicate the 16\% and 84\% quantiles of the inferred posterior distribution of $E_{NN}^0$ while the central value is the mean of the respective posterior.
    }
	\label{fig:EB_reco}
\end{figure}

As shown earlier, the pion component diminishes fairly quickly after the onset of the burst. After three seconds, the bremsstrahlung component dominates. We continue to track the reconstructed value of $E_{NN}^0$ up to five seconds after the core bounce. The results of the inferred values and uncertainties are shown as purple data points in Fig.~\ref{fig:EB_reco}. Already after two seconds, the pionic component sufficiently decreases in intensity to better constrain $E_{NN}^0$. The bi-modal shape of the posterior distribution vanishes. At even later times, we are left with a pure bremsstrahlung component rendering the parameter reconstruction more reliable as implied by the reduced size of the error bars. Therefore, while the reconstruction of $E_{NN}^0$ is technically challenging in the presence of a strong pionic component, it becomes feasible after the decay of this contribution.

\medskip

\begin{figure} [t!]
\centering
    \includegraphics[width=0.8\textwidth]{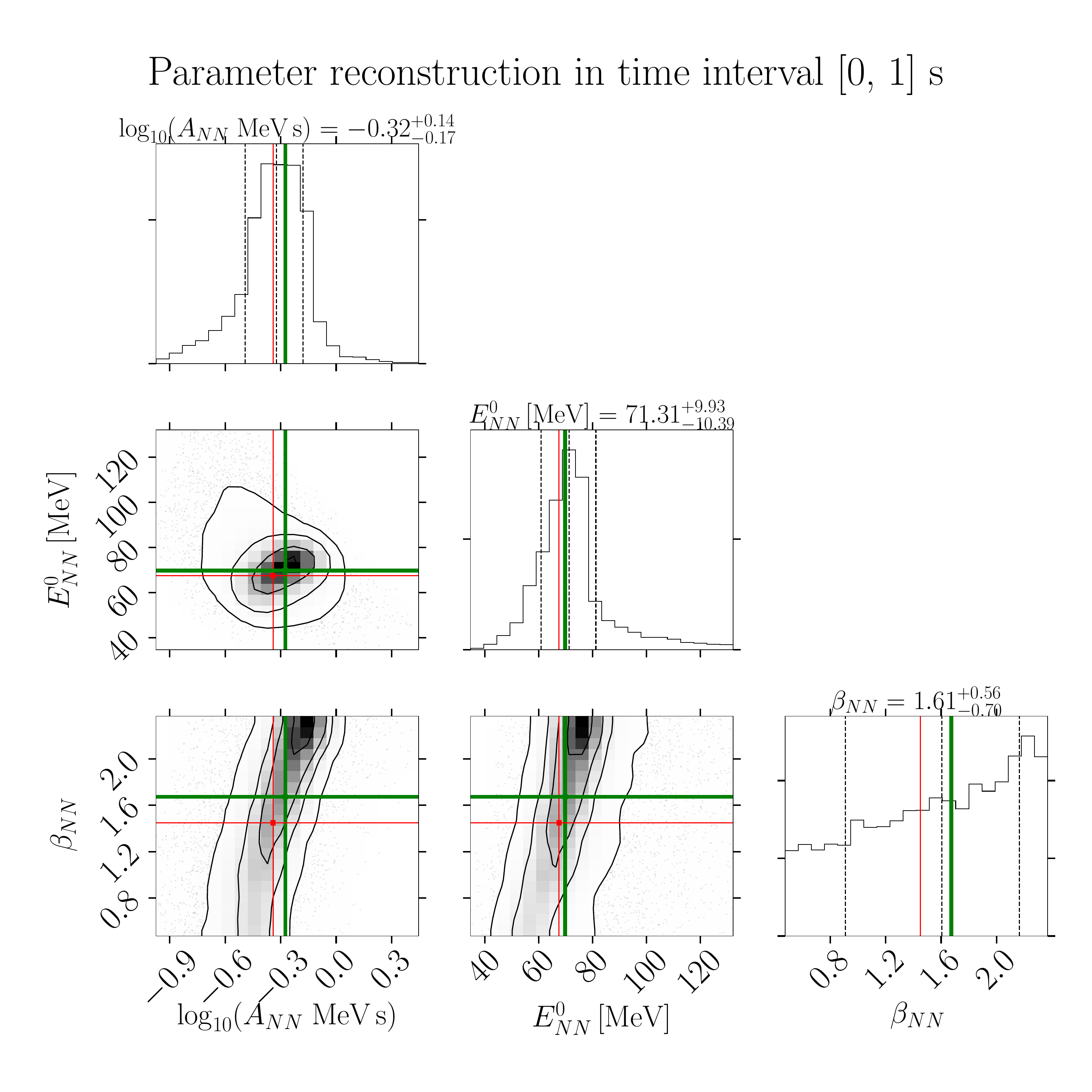}
    \caption{Same as Fig.~\ref{fig:posteriors_paramReco_Brems+pion_dT_1s} assuming a gamma-ray burst fueled only by bremsstrahlung processes.}
	\label{fig:posteriors_paramReco_Brems_dT_1s}
\end{figure}

Finally, as an alternative scenario, we examine a gamma-ray burst simulation considering only the bremsstrahlung contribution that we fit with the respective three-parameter model. Figure~\ref{fig:posteriors_paramReco_Brems_dT_1s} serves as an example of the parameter inference results; again for up to one second after the core bounce. As in a previous study of the LAT capability to study ALPs in Galactic SN explosions \cite{Calore:2023srn}, we find that the spectral index $\beta_{NN}$ of the spectrum is rather unconstrained. The main reason for the difficulties in the inference of $\beta_{NN}$ is the poor energy resolution of the LAT at $\mathcal{O}(100)$ MeV. 

In contrast to the full burst model above, however, the normalization $A_{NN}$ and mean energy $E_{NN}^0$ exhibit a much better reconstruction quality. The time evolution of the latter and its good reconstruction from the data are shown as blue data points in Fig.~\ref{fig:EB_reco}. As anticipated, after three seconds, the inference results of both scenarios are almost identical due to the absence of a pionic contribution. Ultimately, LAT observations of such gamma-ray bursts suffice to constrain $E_{NN}^0$ well enough to estimate the SN core temperature, which we perform in the following Section. 

\subsection{Reconstruction of the SN temperature}
\label{sec:TemperatureReconstruction}
\begin{figure} [t!]
\centering
    \includegraphics[scale=0.6]{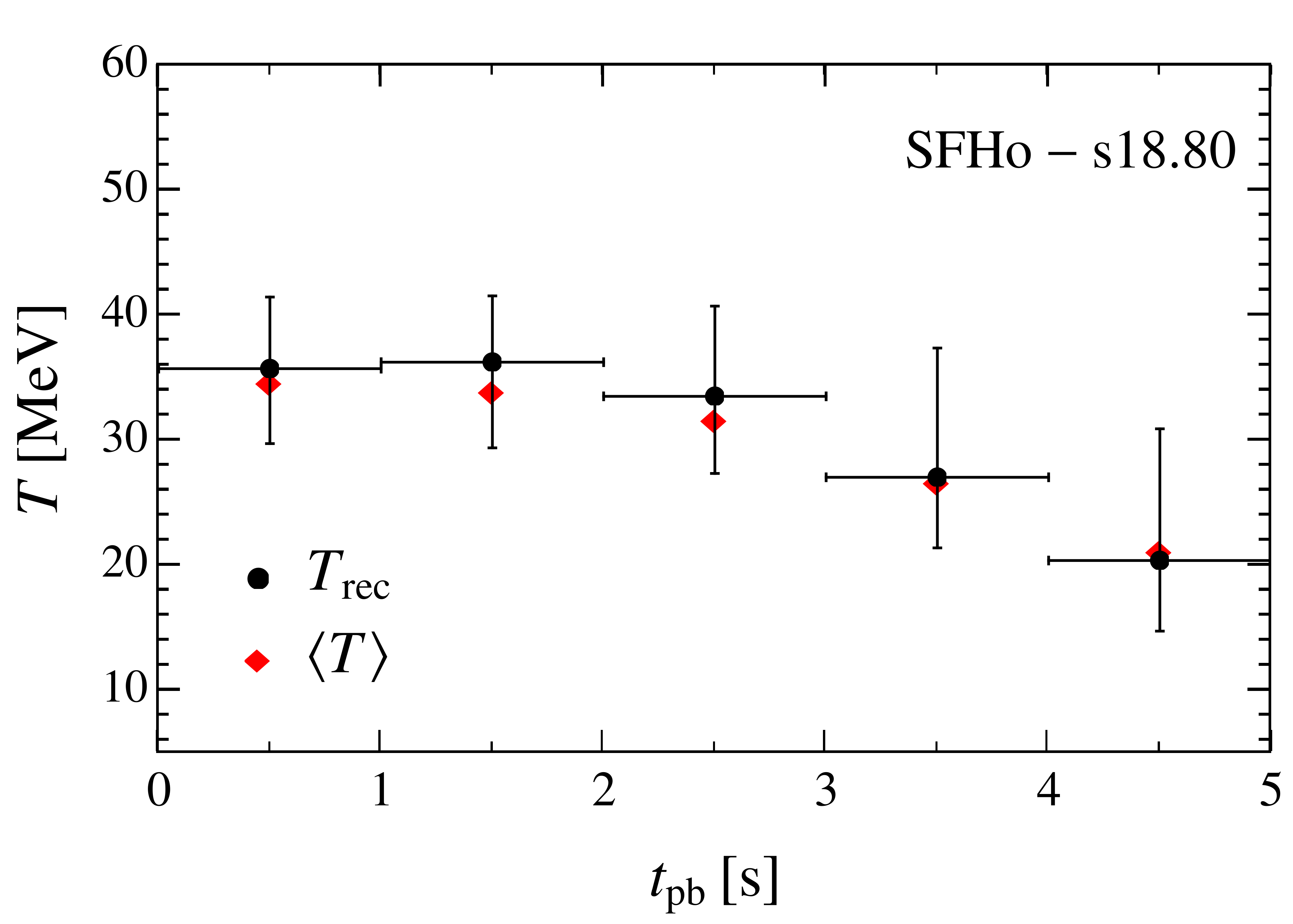}
    \caption{Values of the temperatures reconstructed by employing the procedure illustrated in Sec.~\ref{sec:TemperatureReconstruction} (black markers), together with the average temperatures in the regions of ALP production computed as in Eq.~\eqref{eq:AvgTemperature} (red markers). The values of the temperature have been obtained by removing the systematics due to the temporal fit used within the Fermi Science Tools. In particular, the two panels refer to the first 5 snapshots of our reference SN model SFHo-s18.8.}
	\label{fig:TvsTime_SFHo-s18.8}
\end{figure}

The  estimation of the temperature in the PNS can be performed following the strategy described in Sec.~\ref{subsec:onezone}. For this purpose, we will consider exclusively the energy spectra associated with the $NN$ bremsstrahlung simulated signal only, since we have seen that the reconstruction of parameters in the presence of the pion contribution will not be optimal (and yet possible). Moreover, as discussed in previous sections, the behavior of pions in the SN core is still debated and they may undergo Bose-Einstein condensation in SN simulations predicting high densities in the PNS. Therefore, the study of the pion conversion spectra would not allow for a temperature determination which is truly independent of the other hydrodynamic properties.\\
Starting from the values of $E_{NN}^0$ inferred through the \fermi-LAT analysis illustrated in Sec.~\ref{sec:LAT-ParamReconstruction} and by employing Eq.~(\ref{eq:TRecForm}), one can directly reconstruct the value for the SN temperature in the region relevant for ALP production at $R\sim5-10 \,\  \km$, where the peak in temperature is located.
We highlight that the procedure illustrated in Sec.~\ref{subsec:onezone} is completely unrelated to the realistic SN model considered, since it assumes a simplistic one-zone model where the only free parameter is the temperature.
In Fig.~\ref{fig:TvsTime_SFHo-s18.8} we show the values of the reconstructed temperature $T_{\rm rec}$ of our reference SN model SFHo-s18.8 at $t_{\rm pb}=1-5\,\s$ with their corresponding uncertainty bars. The red markers display the values of the average temperature in the ALP production region $\langle T \rangle$, defined as 
\begin{equation}
    \langle T \rangle=\frac{\int_0^{R_{\rm c}}T(r)\,\frac{d^2N_a}{dVdt}\,dR}{\int_0^{R_{\rm c}}\,\frac{d^2N_a}{dVdt}\,dR}\,,
    \label{eq:AvgTemperature}
\end{equation}
where $R_{\rm c}=15\,\km$ and ${d^2N_a}/{dVdt}$ is the ALP production density. 
Thus, from this definition, it is clear that $\langle T \rangle$ can be employed as an estimator of the average SN temperature in the region $R\sim5-10\,\km$.
We observe that the values of the temperatures reconstructed from the \fermi-LAT results are always 1-$\sigma$ compatible with the corresponding values of $\langle T \rangle$. Therefore, these results suggest that, in case of an observation of the ALP burst from the future Galactic SN, the reconstruction of the $E_{NN}^0$ parameter would immediately provide a good estimation of the SN temperature in the inner core.

\begin{figure} [t!]
\centering
    \includegraphics[width=1\columnwidth]{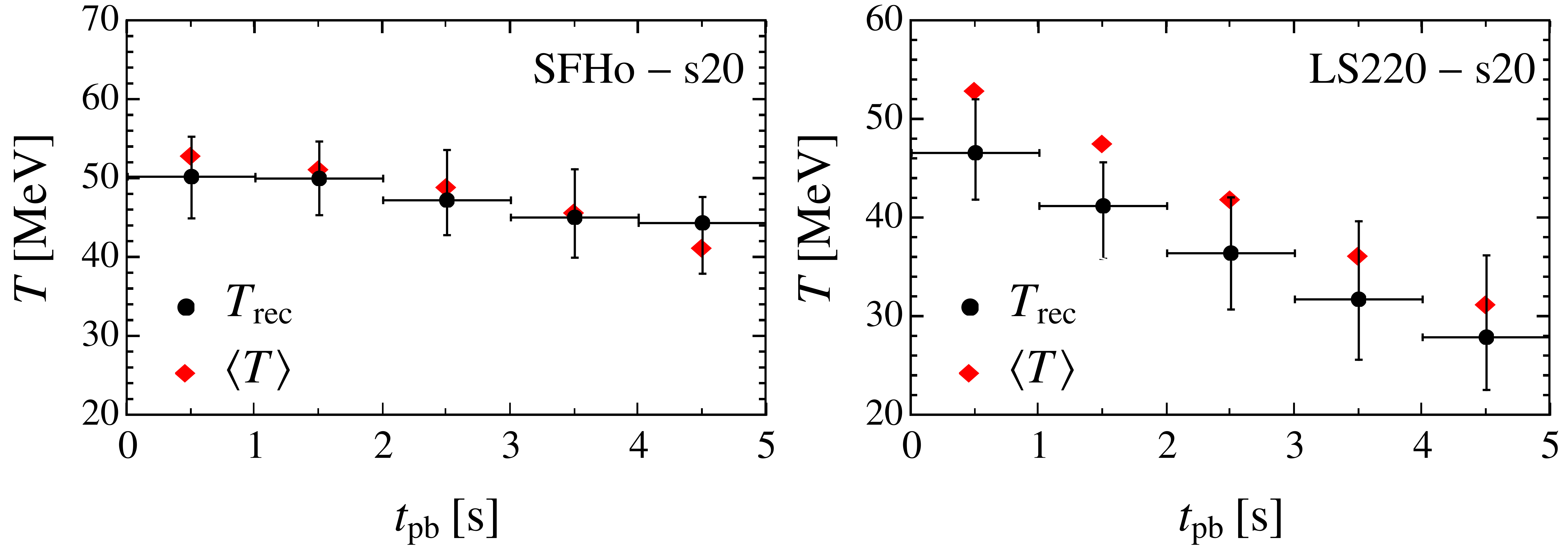}
    \caption{Values of the reconstructed temperature (black markers) for the SFHo-s20 and LS220-s20 SN models and for the first 5 time snapshots considered. Red markers depict the average temperature in the ALP production region.}	\label{fig:Temperature_reconstruction_OtherModels}
\end{figure}

To test the validity of our methodology for temperature reconstruction, we have applied the same procedure illustrated before over the other two different SN models considered in this work. The left panel of Fig.~\ref{fig:Temperature_reconstruction_OtherModels} displays the results for the SFHo-s20 model. As we can appreciate, this model is characterized by higher values of the average PNS temperature, since it leads to a higher PNS mass concerning SFHo-s18.8. Nevertheless, also in this case we can reconstruct the temperature with very high precision, with discrepancies never larger than 1-$\sigma$ in the first $5\,\s$. In the same way, the results obtained for the LS220-s20 SN model are shown in the right panel of Fig.~\ref{fig:Temperature_reconstruction_OtherModels}.
Comparing the outcomes for the SFHo-s20 and LS220-s20 models gives some insights into how this procedure works by keeping fixed the final PNS mass and changing just the EoS employed in the simulation. We notice that the reconstructed values of the temperature for LS220 EoS show a larger discrepancy than observed for SFHo EoS, especially in the first instants after the core bounce. In particular, it seems that the temperature tends to be underestimated. This difference could be related to the behavior of the cooling lightcurves for LS220 EoS which typically appear to be different with respect to the SFHo EoS~\cite{Lucente:2024ngp} which is employed as benchmark in our temperature reconstruction procedure in Sec.~\ref{subsec:onezone}. In this regard we highlight that, as further discussed in Refs.~\cite{Fiorillo:2023frv, Lucente:2024ngp}, DD2, SFHx and SFHo EoS lead to a rather similar time evolution for the SN conditions and neutrino lightcurves, slightly different from LS220. This feature may introduce an additional systematic uncertainty in the analysis. Although these slightly larger deviations, our procedure provides temperature values barely 1-$\sigma$ compatible (and always 2-$\sigma$ compatible) with the real temperature observed from the simulation.\\
Furthermore, it is interesting to observe that the trend of the black points in Fig.~\ref{fig:TvsTime_SFHo-s18.8} and Fig.~\ref{fig:Temperature_reconstruction_OtherModels} reproduces the peculiar behavior of each EoS during the cooling phase. In particular, it is possible to appreciate how, in the first $5\,\s$ after the core bounce, the PNS in the LS220-s20 model cools faster than for the SFHo-s20 model. Therefore, these results suggest that the observation of an ALP burst from the next Galactic SN could also provide some insights about the EoS characterizing the SN core.

\section{Conclusions}
\label{sec:Conclusion}

In this work, we have characterized the physics
case for ultralight ALPs coupled to both nucleons and photons, and emitted by a PNS during a core-collapse SN event.
In particular, we have considered ALP emitted via $NN$ bremsstrahlung, together with a second
contribution from pionic Compton-like scattering off nucleons. We have analyzed the phenomenological implications
of such an emission model and how this signal can be revealed by a temporal and spectral analysis of \fermi-LAT gamma-ray data.

We have shown that, despite the existing SN bounds on the combination of  $g_{ap}$ and 
$g_{a \gamma}$, in the case of a Galactic SN exploding in the field of view of {\it Fermi} LAT we expect
an observable ALP-induced gamma-ray burst, encoding information about the underlying ALP emission mechanism.
Values of $g_{ap}\lesssim5\times10^{-10}$ and $g_{a\gamma} \lesssim 10^{-15}$ GeV$^{-1}$, not accessible in other laboratory or astrophysical experiments, are in the reach of this type of search. 

We developed a Bayesian analysis framework of the simulated ALP-induced gamma-ray signal (fully convolving with 
the instrument response function), demonstrating that we can claim evidence for the pion contribution on top of the
$NN$ one. On the other hand, the parameter reconstruction of the $NN + \pi N$ model is limited by the persistence of large 
degenerancies due to the poor energy resolution and effective area at low energy, depleting the sensitivity to the $NN$ peak, at least in the first second after the burst's onset.
At later times, we demonstrated that the parameter reconstruction becomes more reliable. Therefore, while the reconstruction of $E_{NN}^0$ is technically challenging in the presence of a strong pionic component, it becomes feasible after the decay of this contribution, i.e.~starting from two seconds after the core bounce.

In absence of pion contribution, instead, the \fermi-LAT performance is sufficient to achieve a satisfactory reconstruction 
of the spectral fitting parameters already within the first second after the onset.
In particular, we showed how the inference of the average energy parameter $E_{NN}^0$ is key to providing information on the
average PNS temperature.
In the future, these limitations can be overcome by gamma-ray instruments covering sub-GeV energies (at least down to a few MeV) up to $100$ MeV.
With a field-of-view comparable to the LAT, instruments like e.g.~GRAMS~\cite{Aramaki:2019bpi}, AMEGO-X~\cite{Fleischhack:2021mhc}, e-ASTROGAM~\cite{e-ASTROGAM:2016bph} will provide larger effective areas, a better energy resolution and improved sensitivity at energies relevant for the discrimination between the bremsstrahlung and pion conversion. Therefore, they can play a pivotal role in the detection and characterization of a future core-collapse SN in the Galaxy after the end of the {\it Fermi}-LAT operational period. Finally, the use of CubeSats (like the funded COSI~\cite{Tomsick:2021wed}) can significantly increase the sky coverage, at the expenses of flux sensitivity though.

We recall that, since ALPs are produced in the inner core of a PNS, their detection will allow us to probe stellar regions where neutrinos are no viable messengers, as they are emitted at larger radii.
Therefore, a multi-messenger SN detection exploiting both ALPs and neutrinos can provide us with unique access to the fundamental processes and conditions regulating SN physics. 

\section*{Acknowledgements}

We warmly thank  Thomas Janka for giving us access to the {\tt GARCHING} group archive. 
AL kindly thanks LAPTh for their hospitality during its visit.
This article is based upon work from COST Action COSMIC WISPers CA21106, supported by COST (European Cooperation in Science and Technology).
The work of PC is supported by the European Research Council under Grant No.~742104 and by the Swedish Research Council (VR) under grants 2018-03641 and 2019-02337. The work of CE is supported by the ANR through grant ANR-19-CE31-0005-01 (PI: F.~Calore), and has been supported by the EOSC Future project which is co-funded by the European Union Horizon Programme call INFRAEOSC-03-2020, Grant Agreement 101017536. This publication is supported by the European Union's Horizon Europe research and innovation programme under the Marie Sk\l odowska-Curie Postdoctoral Fellowship Programme, SMASH co-funded under the grant agreement No.~101081355.
The work of AM and AL was partially supported by the research grant number 2022E2J4RK ``PANTHEON: Perspectives in Astroparticle and
Neutrino THEory with Old and New messengers" under the program PRIN 2022 funded by the Italian Ministero dell’Universit\`a e della Ricerca (MUR). GL is supported by the European Union’s Horizon 2020 Europe research and innovation programme under the Marie Skłodowska-Curie grant agreement No 860881-HIDDeN.
This work is (partially) supported
by ICSC – Centro Nazionale di Ricerca in High Performance Computing.

\bibliographystyle{bibi.bst}
\bibliography{references.bib}

\providecommand{\href}[2]{#2}\begingroup\raggedright\begin{thebibliography}{10}

\bibitem{Caputo:2024oqc}
A.~Caputo and G.~Raffelt, \emph{{Astrophysical Axion Bounds: The 2024 Edition}}, \href{https://doi.org/10.22323/1.454.0041}{\emph{PoS} {\bfseries COSMICWISPers} (2024) 041} [\href{https://arxiv.org/abs/2401.13728}{{\ttfamily 2401.13728}}].

\bibitem{Lella:2022uwi}
A.~Lella, P.~Carenza, G.~Lucente, M.~Giannotti and A.~Mirizzi, \emph{{Protoneutron stars as cosmic factories for massive axionlike particles}}, \href{https://doi.org/10.1103/PhysRevD.107.103017}{\emph{Phys. Rev. D} {\bfseries 107} (2023) 103017} [\href{https://arxiv.org/abs/2211.13760}{{\ttfamily 2211.13760}}].

\bibitem{Lella:2023bfb}
A.~Lella, P.~Carenza, G.~Co', G.~Lucente, M.~Giannotti, A.~Mirizzi and T.~Rauscher, \emph{{Getting the most on supernova axions}}, \href{https://doi.org/10.1103/PhysRevD.109.023001}{\emph{Phys. Rev. D} {\bfseries 109} (2024) 023001} [\href{https://arxiv.org/abs/2306.01048}{{\ttfamily 2306.01048}}].

\bibitem{Payez:2014xsa}
A.~Payez, C.~Evoli, T.~Fischer, M.~Giannotti, A.~Mirizzi and A.~Ringwald, \emph{{Revisiting the SN1987A gamma-ray limit on ultralight axion-like particles}}, \href{https://doi.org/10.1088/1475-7516/2015/02/006}{\emph{JCAP} {\bfseries 02} (2015) 006} [\href{https://arxiv.org/abs/1410.3747}{{\ttfamily 1410.3747}}].

\bibitem{Meyer:2016wrm}
M.~Meyer, M.~Giannotti, A.~Mirizzi, J.~Conrad and M.~A. S\'anchez-Conde, \emph{{Fermi Large Area Telescope as a Galactic Supernovae Axionscope}}, \href{https://doi.org/10.1103/PhysRevLett.118.011103}{\emph{Phys. Rev. Lett.} {\bfseries 118} (2017) 011103} [\href{https://arxiv.org/abs/1609.02350}{{\ttfamily 1609.02350}}].

\bibitem{Calore:2023srn}
F.~Calore, P.~Carenza, C.~Eckner, M.~Giannotti, G.~Lucente, A.~Mirizzi and F.~Sivo, \emph{{Uncovering axionlike particles in supernova gamma-ray spectra}}, \href{https://doi.org/10.1103/PhysRevD.109.043010}{\emph{Phys. Rev. D} {\bfseries 109} (2024) 043010} [\href{https://arxiv.org/abs/2306.03925}{{\ttfamily 2306.03925}}].

\bibitem{Calore:2020tjw}
F.~Calore, P.~Carenza, M.~Giannotti, J.~Jaeckel and A.~Mirizzi, \emph{{Bounds on axionlike particles from the diffuse supernova flux}}, \href{https://doi.org/10.1103/PhysRevD.102.123005}{\emph{Phys. Rev. D} {\bfseries 102} (2020) 123005} [\href{https://arxiv.org/abs/2008.11741}{{\ttfamily 2008.11741}}].

\bibitem{Muller:2023pip}
E.~M\"uller, P.~Carenza, C.~Eckner and A.~Goobar, \emph{{Constraining MeV-scale axionlike particles with Fermi-LAT observations of SN 2023ixf}}, \href{https://doi.org/10.1103/PhysRevD.109.023018}{\emph{Phys. Rev. D} {\bfseries 109} (2024) 023018} [\href{https://arxiv.org/abs/2306.16397}{{\ttfamily 2306.16397}}].

\bibitem{Mirizzi:2015eza}
A.~Mirizzi, I.~Tamborra, H.-T. Janka, N.~Saviano, K.~Scholberg, R.~Bollig, L.~Hudepohl and S.~Chakraborty, \emph{{Supernova Neutrinos: Production, Oscillations and Detection}}, \href{https://doi.org/10.1393/ncr/i2016-10120-8}{\emph{Riv. Nuovo Cim.} {\bfseries 39} (2016) 1} [\href{https://arxiv.org/abs/1508.00785}{{\ttfamily 1508.00785}}].

\bibitem{Carena:1988kr}
M.~Carena and R.~D. Peccei, \emph{{The Effective Lagrangian for Axion Emission From {SN1987A}}}, \href{https://doi.org/10.1103/PhysRevD.40.652}{\emph{Phys. Rev. D} {\bfseries 40} (1989) 652}.

\bibitem{Brinkmann:1988vi}
R.~P. Brinkmann and M.~S. Turner, \emph{{Numerical Rates for Nucleon-Nucleon Axion Bremsstrahlung}}, \href{https://doi.org/10.1103/PhysRevD.38.2338}{\emph{Phys. Rev. D} {\bfseries 38} (1988) 2338}.

\bibitem{Raffelt:1993ix}
G.~Raffelt and D.~Seckel, \emph{{A selfconsistent approach to neutral current processes in supernova cores}}, \href{https://doi.org/10.1103/PhysRevD.52.1780}{\emph{Phys. Rev. D} {\bfseries 52} (1995) 1780} [\href{https://arxiv.org/abs/astro-ph/9312019}{{\ttfamily astro-ph/9312019}}].

\bibitem{Raffelt:1996wa}
G.~G. Raffelt, \emph{{Stars as laboratories for fundamental physics}: {The astrophysics of neutrinos, axions, and other weakly interacting particles}}. 5, 1996.

\bibitem{Turner:1991ax}
M.~S. Turner, \emph{{Dirac neutrinos and SN1987A}}, \href{https://doi.org/10.1103/PhysRevD.45.1066}{\emph{Phys. Rev. D} {\bfseries 45} (1992) 1066}.

\bibitem{Keil:1996ju}
W.~Keil, H.-T. Janka, D.~N. Schramm, G.~Sigl, M.~S. Turner and J.~R. Ellis, \emph{{A Fresh look at axions and SN-1987A}}, \href{https://doi.org/10.1103/PhysRevD.56.2419}{\emph{Phys. Rev. D} {\bfseries 56} (1997) 2419} [\href{https://arxiv.org/abs/astro-ph/9612222}{{\ttfamily astro-ph/9612222}}].

\bibitem{Carenza:2019pxu}
P.~Carenza, T.~Fischer, M.~Giannotti, G.~Guo, G.~Mart\'\i{}nez-Pinedo and A.~Mirizzi, \emph{{Improved axion emissivity from a supernova via nucleon-nucleon bremsstrahlung}}, \href{https://doi.org/10.1088/1475-7516/2019/10/016}{\emph{JCAP} {\bfseries 10} (2019) 016} [\href{https://arxiv.org/abs/1906.11844}{{\ttfamily 1906.11844}}]. [Erratum: JCAP 05, E01 (2020)].

\bibitem{OHare:2020wum}
C.~A.~J. O'Hare, A.~Caputo, A.~J. Millar and E.~Vitagliano, \emph{{Axion helioscopes as solar magnetometers}}, \href{https://doi.org/10.1103/PhysRevD.102.043019}{\emph{Phys. Rev. D} {\bfseries 102} (2020) 043019} [\href{https://arxiv.org/abs/2006.10415}{{\ttfamily 2006.10415}}].

\bibitem{Hoof:2023jol}
S.~Hoof, J.~Jaeckel and L.~J. Thormaehlen, \emph{{Axion helioscopes as solar thermometers}}, \href{https://doi.org/10.1088/1475-7516/2023/10/024}{\emph{JCAP} {\bfseries 10} (2023) 024} [\href{https://arxiv.org/abs/2306.00077}{{\ttfamily 2306.00077}}].

\bibitem{Fore:2019wib}
B.~Fore and S.~Reddy, \emph{{Pions in hot dense matter and their astrophysical implications}}, \href{https://doi.org/10.1103/PhysRevC.101.035809}{\emph{Phys. Rev. C} {\bfseries 101} (2020) 035809} [\href{https://arxiv.org/abs/1911.02632}{{\ttfamily 1911.02632}}].

\bibitem{Fore:2023gwv}
B.~Fore, N.~Kaiser, S.~Reddy and N.~C. Warrington, \emph{{The mass of charged pions in neutron star matter}},  \href{https://arxiv.org/abs/2301.07226}{{\ttfamily 2301.07226}}.

\bibitem{Carenza:2020cis}
P.~Carenza, B.~Fore, M.~Giannotti, A.~Mirizzi and S.~Reddy, \emph{{Enhanced Supernova Axion Emission and its Implications}}, \href{https://doi.org/10.1103/PhysRevLett.126.071102}{\emph{Phys. Rev. Lett.} {\bfseries 126} (2021) 071102} [\href{https://arxiv.org/abs/2010.02943}{{\ttfamily 2010.02943}}].

\bibitem{Carenza:2023lci}
P.~Carenza, \emph{{Axion emission from supernovae: a cheatsheet}}, \href{https://doi.org/10.1140/epjp/s13360-023-04484-2}{\emph{Eur. Phys. J. Plus} {\bfseries 138} (2023) 836} [\href{https://arxiv.org/abs/2309.14798}{{\ttfamily 2309.14798}}].

\bibitem{Stoica:2009zh}
S.~Stoica, B.~Pastrav, J.~E. Horvath and M.~P. Allen, \emph{{Pion mass effects on axion emission from neutron stars through NN bremsstrahlung processes}}, \href{https://doi.org/10.1016/j.nuclphysa.2009.07.007}{\emph{Nucl. Phys. A} {\bfseries 828} (2009) 439} [\href{https://arxiv.org/abs/0906.3134}{{\ttfamily 0906.3134}}]. [Erratum: Nucl.Phys.A 832, 148 (2010)].

\bibitem{Ericson:1988wr}
T.~E.~O. Ericson and J.~F. Mathiot, \emph{{Axion Emission from SN 1987a: Nuclear Physics Constraints}}, \href{https://doi.org/10.1016/0370-2693(89)91103-9}{\emph{Phys. Lett. B} {\bfseries 219} (1989) 507}.

\bibitem{Raffelt:1991pw}
G.~Raffelt and D.~Seckel, \emph{{Multiple scattering suppression of the bremsstrahlung emission of neutrinos and axions in supernovae}}, \href{https://doi.org/10.1103/PhysRevLett.67.2605}{\emph{Phys. Rev. Lett.} {\bfseries 67} (1991) 2605}.

\bibitem{Janka:1995ir}
H.-T. Janka, W.~Keil, G.~Raffelt and D.~Seckel, \emph{{Nucleon spin fluctuations and the supernova emission of neutrinos and axions}}, \href{https://doi.org/10.1103/PhysRevLett.76.2621}{\emph{Phys. Rev. Lett.} {\bfseries 76} (1996) 2621} [\href{https://arxiv.org/abs/astro-ph/9507023}{{\ttfamily astro-ph/9507023}}].

\bibitem{Choi:2021ign}
K.~Choi, H.~J. Kim, H.~Seong and C.~S. Shin, \emph{{Axion emission from supernova with axion-pion-nucleon contact interaction}}, \href{https://doi.org/10.1007/JHEP02(2022)143}{\emph{JHEP} {\bfseries 02} (2022) 143} [\href{https://arxiv.org/abs/2110.01972}{{\ttfamily 2110.01972}}].

\bibitem{Ho:2022oaw}
S.-Y. Ho, J.~Kim, P.~Ko and J.-h. Park, \emph{{Supernova axion emissivity with \ensuremath{\Delta}(1232) resonance in heavy baryon chiral perturbation theory}}, \href{https://doi.org/10.1103/PhysRevD.107.075002}{\emph{Phys. Rev. D} {\bfseries 107} (2023) 075002} [\href{https://arxiv.org/abs/2212.01155}{{\ttfamily 2212.01155}}].

\bibitem{SNarchive}
\emph{{Garching core-collapse supernova research archive}},  \url{https://wwwmpa.mpa-garching.mpg.de/ccsnarchive//}.

\bibitem{Rampp:2002bq}
M.~Rampp and H.~T. Janka, \emph{{Radiation hydrodynamics with neutrinos: Variable Eddington factor method for core collapse supernova simulations}}, \href{https://doi.org/10.1051/0004-6361:20021398}{\emph{Astron. Astrophys.} {\bfseries 396} (2002) 361} [\href{https://arxiv.org/abs/astro-ph/0203101}{{\ttfamily astro-ph/0203101}}].

\bibitem{Hempel:2009mc}
M.~Hempel and J.~Schaffner-Bielich, \emph{{Statistical Model for a Complete Supernova Equation of State}}, \href{https://doi.org/10.1016/j.nuclphysa.2010.02.010}{\emph{Nucl. Phys. A} {\bfseries 837} (2010) 210} [\href{https://arxiv.org/abs/0911.4073}{{\ttfamily 0911.4073}}].

\bibitem{Steiner:2012rk}
A.~W. Steiner, M.~Hempel and T.~Fischer, \emph{{Core-collapse supernova equations of state based on neutron star observations}}, \href{https://doi.org/10.1088/0004-637X/774/1/17}{\emph{Astrophys. J.} {\bfseries 774} (2013) 17} [\href{https://arxiv.org/abs/1207.2184}{{\ttfamily 1207.2184}}].

\bibitem{Sukhbold:2017cnt}
T.~Sukhbold, S.~Woosley and A.~Heger, \emph{{A High-resolution Study of Presupernova Core Structure}}, \href{https://doi.org/10.3847/1538-4357/aac2da}{\emph{Astrophys. J.} {\bfseries 860} (2018) 93} [\href{https://arxiv.org/abs/1710.03243}{{\ttfamily 1710.03243}}].

\bibitem{Fischer:2021jfm}
T.~Fischer, P.~Carenza, B.~Fore, M.~Giannotti, A.~Mirizzi and S.~Reddy, \emph{{Observable signatures of enhanced axion emission from protoneutron stars}}, \href{https://doi.org/10.1103/PhysRevD.104.103012}{\emph{Phys. Rev. D} {\bfseries 104} (2021) 103012} [\href{https://arxiv.org/abs/2108.13726}{{\ttfamily 2108.13726}}].

\bibitem{Migdal:1990vm}
A.~B. Migdal, E.~E. Saperstein, M.~A. Troitsky and D.~N. Voskresensky, \emph{{Pion degrees of freedom in nuclear matter}}, \href{https://doi.org/10.1016/0370-1573(90)90132-L}{\emph{Phys. Rept.} {\bfseries 192} (1990) 179}.

\bibitem{Woosley:2007as}
S.~E. Woosley and A.~Heger, \emph{{Nucleosynthesis and Remnants in Massive Stars of Solar Metallicity}}, \href{https://doi.org/10.1016/j.physrep.2007.02.009}{\emph{Phys. Rept.} {\bfseries 442} (2007) 269} [\href{https://arxiv.org/abs/astro-ph/0702176}{{\ttfamily astro-ph/0702176}}].

\bibitem{Lattimer:1991nc}
J.~M. Lattimer and F.~D. Swesty, \emph{{A Generalized equation of state for hot, dense matter}}, \href{https://doi.org/10.1016/0375-9474(91)90452-C}{\emph{Nucl. Phys. A} {\bfseries 535} (1991) 331}.

\bibitem{Roberts:2011yw}
L.~F. Roberts, G.~Shen, V.~Cirigliano, J.~A. Pons, S.~Reddy and S.~E. Woosley, \emph{{Proto-Neutron Star Cooling with Convection: The Effect of the Symmetry Energy}}, \href{https://doi.org/10.1103/PhysRevLett.108.061103}{\emph{Phys. Rev. Lett.} {\bfseries 108} (2012) 061103} [\href{https://arxiv.org/abs/1112.0335}{{\ttfamily 1112.0335}}].

\bibitem{GrillidiCortona:2015jxo}
G.~Grilli~di Cortona, E.~Hardy, J.~Pardo~Vega and G.~Villadoro, \emph{{The QCD axion, precisely}}, \href{https://doi.org/10.1007/JHEP01(2016)034}{\emph{JHEP} {\bfseries 01} (2016) 034} [\href{https://arxiv.org/abs/1511.02867}{{\ttfamily 1511.02867}}].

\bibitem{Janka:2006fh}
H.-T. Janka, K.~Langanke, A.~Marek, G.~Martinez-Pinedo and B.~Mueller, \emph{{Theory of Core-Collapse Supernovae}}, \href{https://doi.org/10.1016/j.physrep.2007.02.002}{\emph{Phys. Rept.} {\bfseries 442} (2007) 38} [\href{https://arxiv.org/abs/astro-ph/0612072}{{\ttfamily astro-ph/0612072}}].

\bibitem{GalloRosso:2017mdz}
A.~Gallo~Rosso, F.~Vissani and M.~C. Volpe, \emph{{What can we learn on supernova neutrino spectra with water Cherenkov detectors?}}, \href{https://doi.org/10.1088/1475-7516/2018/04/040}{\emph{JCAP} {\bfseries 04} (2018) 040} [\href{https://arxiv.org/abs/1712.05584}{{\ttfamily 1712.05584}}].

\bibitem{Lujan-Peschard:2014lta}
C.~Lujan-Peschard, G.~Pagliaroli and F.~Vissani, \emph{{Spectrum of Supernova Neutrinos in Ultra-pure Scintillators}}, \href{https://doi.org/10.1088/1475-7516/2014/07/051}{\emph{JCAP} {\bfseries 07} (2014) 051} [\href{https://arxiv.org/abs/1402.6953}{{\ttfamily 1402.6953}}].

\bibitem{GalloRosso:2017hbp}
A.~Gallo~Rosso, F.~Vissani and M.~C. Volpe, \emph{{Measuring the neutron star compactness and binding energy with supernova neutrinos}}, \href{https://doi.org/10.1088/1475-7516/2017/11/036}{\emph{JCAP} {\bfseries 11} (2017) 036} [\href{https://arxiv.org/abs/1708.00760}{{\ttfamily 1708.00760}}].

\bibitem{DeAngelis:2011id}
A.~De~Angelis, G.~Galanti and M.~Roncadelli, \emph{{Relevance of axion-like particles for very-high-energy astrophysics}}, \href{https://doi.org/10.1103/PhysRevD.84.105030}{\emph{Phys. Rev. D} {\bfseries 84} (2011) 105030} [\href{https://arxiv.org/abs/1106.1132}{{\ttfamily 1106.1132}}]. [Erratum: Phys.Rev.D 87, 109903 (2013)].

\bibitem{vanLeeuwen:2007tv}
F.~van Leeuwen, \emph{{Validation of the new Hipparcos reduction}}, \href{https://doi.org/10.1051/0004-6361:20078357}{\emph{Astron. Astrophys.} {\bfseries 474} (2007) 653} [\href{https://arxiv.org/abs/0708.1752}{{\ttfamily 0708.1752}}].

\bibitem{Jansson:2012pc}
R.~Jansson and G.~R. Farrar, \emph{{A New Model of the Galactic Magnetic Field}}, \href{https://doi.org/10.1088/0004-637X/757/1/14}{\emph{Astrophys. J.} {\bfseries 757} (2012) 14} [\href{https://arxiv.org/abs/1204.3662}{{\ttfamily 1204.3662}}].

\bibitem{Planck:2016gdp}
{\scshape Planck} Collaboration, R.~Adam et~al., \emph{{Planck intermediate results.}: {XLII. Large-scale Galactic magnetic fields}}, \href{https://doi.org/10.1051/0004-6361/201528033}{\emph{Astron. Astrophys.} {\bfseries 596} (2016) A103} [\href{https://arxiv.org/abs/1601.00546}{{\ttfamily 1601.00546}}].

\bibitem{Pshirkov:2011um}
M.~S. Pshirkov, P.~G. Tinyakov, P.~P. Kronberg and K.~J. Newton-McGee, \emph{{Deriving global structure of the Galactic Magnetic Field from Faraday Rotation Measures of extragalactic sources}}, \href{https://doi.org/10.1088/0004-637X/738/2/192}{\emph{Astrophys. J.} {\bfseries 738} (2011) 192} [\href{https://arxiv.org/abs/1103.0814}{{\ttfamily 1103.0814}}].

\bibitem{Calore:2021hhn}
F.~Calore, P.~Carenza, C.~Eckner, T.~Fischer, M.~Giannotti, J.~Jaeckel, K.~Kotake, T.~Kuroda, A.~Mirizzi and F.~Sivo, \emph{{3D template-based Fermi-LAT constraints on the diffuse supernova axion-like particle background}}, \href{https://doi.org/10.1103/PhysRevD.105.063028}{\emph{Phys. Rev. D} {\bfseries 105} (2022) 063028} [\href{https://arxiv.org/abs/2110.03679}{{\ttfamily 2110.03679}}].

\bibitem{Healy:2023ovi}
S.~Healy, S.~Horiuchi, M.~Colomer~Molla, D.~Milisavljevic, J.~Tseng, F.~Bergin, K.~Weil and M.~Tanaka, \emph{{Red Supergiant Candidates for Multimessenger Monitoring of the Next Galactic Supernova}}, \href{https://doi.org/10.1093/mnras/stae738}{\emph{Mon. Not. Roy. Astron. Soc.} {\bfseries 529} (2024) 3630} [\href{https://arxiv.org/abs/2307.08785}{{\ttfamily 2307.08785}}].

\bibitem{Feroz:2008xx}
F.~Feroz, M.~P. Hobson and M.~Bridges, \emph{{MultiNest: an efficient and robust Bayesian inference tool for cosmology and particle physics}}, \href{https://doi.org/10.1111/j.1365-2966.2009.14548.x}{\emph{Mon. Not. Roy. Astron. Soc.} {\bfseries 398} (2009) 1601} [\href{https://arxiv.org/abs/0809.3437}{{\ttfamily 0809.3437}}].

\bibitem{Ning:2024eky}
O.~Ning and B.~R. Safdi, \emph{{Leading Axion-Photon Sensitivity with NuSTAR Observations of M82 and M87}},  \href{https://arxiv.org/abs/2404.14476}{{\ttfamily 2404.14476}}.

\bibitem{Trotta:2008qt}
R.~Trotta, \emph{Bayes in the sky: {{Bayesian}} inference and model selection in cosmology}, \href{https://doi.org/10.1080/00107510802066753}{\emph{Contemp. Phys.} {\bfseries 49} (2008) 71} [\href{https://arxiv.org/abs/0803.4089}{{\ttfamily 0803.4089}}].

\bibitem{Jaeckel:2017tud}
J.~Jaeckel, P.~C. Malta and J.~Redondo, \emph{{Decay photons from the axionlike particles burst of type II supernovae}}, \href{https://doi.org/10.1103/PhysRevD.98.055032}{\emph{Phys. Rev. D} {\bfseries 98} (2018) 055032} [\href{https://arxiv.org/abs/1702.02964}{{\ttfamily 1702.02964}}].

\bibitem{Hoof:2022xbe}
S.~Hoof and L.~Schulz, \emph{{Updated constraints on axion-like particles from temporal information in supernova SN1987A gamma-ray data}}, \href{https://doi.org/10.1088/1475-7516/2023/03/054}{\emph{JCAP} {\bfseries 03} (2023) 054} [\href{https://arxiv.org/abs/2212.09764}{{\ttfamily 2212.09764}}].

\bibitem{Muller:2023vjm}
E.~M\"uller, F.~Calore, P.~Carenza, C.~Eckner and M.~C.~D. Marsh, \emph{{Investigating the gamma-ray burst from decaying MeV-scale axion-like particles produced in supernova explosions}}, \href{https://doi.org/10.1088/1475-7516/2023/07/056}{\emph{JCAP} {\bfseries 07} (2023) 056} [\href{https://arxiv.org/abs/2304.01060}{{\ttfamily 2304.01060}}].

\bibitem{Lucente:2024ngp}
G.~Lucente, M.~Heinlein, H.~T. Janka and A.~Mirizzi, \emph{{Simple fits for the neutrino luminosities from protoneutron star cooling}},  \href{https://arxiv.org/abs/2405.00769}{{\ttfamily 2405.00769}}.

\bibitem{Fiorillo:2023frv}
D.~F.~G. Fiorillo, M.~Heinlein, H.-T. Janka, G.~Raffelt, E.~Vitagliano and R.~Bollig, \emph{{Supernova simulations confront SN 1987A neutrinos}}, \href{https://doi.org/10.1103/PhysRevD.108.083040}{\emph{Phys. Rev. D} {\bfseries 108} (2023) 083040} [\href{https://arxiv.org/abs/2308.01403}{{\ttfamily 2308.01403}}].

\bibitem{Aramaki:2019bpi}
T.~Aramaki, P.~Hansson~Adrian, G.~Karagiorgi and H.~Odaka, \emph{{Dual MeV Gamma-Ray and Dark Matter Observatory - GRAMS Project}}, \href{https://doi.org/10.1016/j.astropartphys.2019.07.002}{\emph{Astropart. Phys.} {\bfseries 114} (2020) 107} [\href{https://arxiv.org/abs/1901.03430}{{\ttfamily 1901.03430}}].

\bibitem{Fleischhack:2021mhc}
H.~Fleischhack, \emph{{AMEGO-X: MeV gamma-ray Astronomy in the Multi-messenger Era}}, \href{https://doi.org/10.22323/1.395.0649}{\emph{PoS} {\bfseries ICRC2021} (2021) 649} [\href{https://arxiv.org/abs/2108.02860}{{\ttfamily 2108.02860}}].

\bibitem{e-ASTROGAM:2016bph}
{\scshape e-ASTROGAM} Collaboration, A.~De~Angelis et~al., \emph{{The e-ASTROGAM mission}}, \href{https://doi.org/10.1007/s10686-017-9533-6}{\emph{Exper. Astron.} {\bfseries 44} (2017) 25} [\href{https://arxiv.org/abs/1611.02232}{{\ttfamily 1611.02232}}].

\bibitem{Tomsick:2021wed}
{\scshape COSI} Collaboration, J.~A. Tomsick, \emph{{The Compton Spectrometer and Imager Project for MeV Astronomy}}, \href{https://doi.org/10.22323/1.395.0652}{\emph{PoS} {\bfseries ICRC2021} (2021) 652} [\href{https://arxiv.org/abs/2109.10403}{{\ttfamily 2109.10403}}].

\end{thebibliography}\endgroup

\clearpage

\setcounter{equation}{0}
\setcounter{figure}{0}
\setcounter{table}{0}
\setcounter{section}{0}
\makeatletter
\renewcommand{\theequation}{A\arabic{equation}}
\renewcommand{\thefigure}{A\arabic{figure}}
\renewcommand{\thetable}{A\arabic{table}}

\appendix
\section{SN model fitting parameters}
\label{app:Parameters}
In this Appendix we show the values of the fitting parameters employed to compute the ALP production spectra for the different SN models we used, computed for $g_{ap}=5\times 10^{-10}$ and $g_{an}=0$ in the time interval $t_{\rm pb} \in [1,\,8]~\s$.
\subsection{SFHo-s18.8}
\label{app:SFHo-s18.8}
In Tab.~\ref{tab:FittingParam} we provide the values of the fitting parameters employed in Eq.~\eqref{eq:BremFit} and Eq.~\eqref{eq:PionFit} for our benchmark SN model SFHo-s18.8, described in Sec.~\ref{sec:ALP production}. To extract the time-dependence of the ALP spectrum, the parameters in Tab.~\ref{tab:FittingParam} can be well interpolated by means of the following functional forms
\begin{equation}
    \begin{split}
        E^0_{NN}&=\Tilde{E}^0_{NN}\,t_{\mathrm{s}}^{0.755}\,e^{-0.413\,t_{\mathrm{s}}}\,,\\
        \beta_{NN}&=\Tilde{\beta}_{NN}\,t_{\mathrm{s}}^{0.0410}\,e^{-0.0542\,t_{\mathrm{s}}}\,,\\
        A_{NN}&=\Tilde{A}_{NN}\,t_{\mathrm{s}}^{1.865}\,e^{-1.345\,t_{\mathrm{s}}}\,,\\
    \end{split}
\end{equation}

\begin{equation}
    \begin{split}
        E^0_{\pi N}&=\Tilde{E}^0_{\pi N}\,t_{\mathrm{s}}^{0.304}\,e^{-0.542\,t_{\mathrm{s}}}\,,\\
        \beta_{\pi N}&=\Tilde{\beta}_{\pi N}\,t_{\mathrm{s}}^{-0.503}\,e^{-0.019\,t_{\mathrm{s}}}\,,\\
        A_{\pi N}&=\Tilde{A}_{\pi N}\,t_{\mathrm{s}}^{5.975}\,e^{-4.944\,t_{\mathrm{s}}}\,,\\
        \omega_{\pi N}&=\Tilde{\omega}_{\pi N}\,(1+1.537\,t_{\mathrm{s}}^{0.050})\,,\,
    \end{split}
\end{equation}
where $t_{\s}=t_{\mathrm{pb}}/1\,\s$. In particular, for $NN$ bremsstrahlung ${\Tilde{E}^0_{NN}=102.10\,\MeV}$, ${\Tilde{\beta}_{NN}=1.53}$, ${\Tilde{A}_{NN}=1.75\times10^{55}\,\MeV^{-1}\,\s^{-1}}$, while for pion conversion
$\Tilde{E}^0_{\pi N}=218.59\,\MeV$, $\Tilde{\beta}_{\pi N}=1.27$, $\Tilde{A}_{\pi N}=3.88\times10^{56}\,\MeV^{-1}\,\s^{-1}$ and $\Tilde{\omega}_c=40.07\,\MeV$.

\subsection{SFHo-s20}
\label{app:SFHo-s20}
In Tab.~\ref{tab:FittingParam2} we provide the values of the fitting parameters employed in Eq.~\eqref{eq:BremFit} for th SN model SFHo-s20, described in Sec.~\ref{sec:TemperatureReconstruction}.
The time dependence for fitting parameters for this SN model can be extracted as
\begin{equation}
    \begin{split}
        E^0_{NN}&=\Tilde{E}^0_{NN}\,t_{\mathrm{s}}^{0.368}\,e^{-0.192\,t_{\mathrm{s}}}\,,\\
        \beta_{NN}&=\Tilde{\beta}_{NN}\,t_{\mathrm{s}}^{0.0531}\,e^{-0.0420\,t_{\mathrm{s}}}\,,\\
        A_{NN}&=\Tilde{A}_{NN}\,t_{\mathrm{s}}^{1.22}\,e^{-0.768\,t_{\mathrm{s}}}\,,\\
    \end{split}
\end{equation}
where $\Tilde{E}^0_{NN}=101.29\,\MeV$, $\Tilde{\beta}_{NN}=1.53$ , $\Tilde{A}_{NN}=4.13\times10^{55}\,\MeV^{-1}\,\s^{-1}$.\\


\subsection{LS220-s20}
\label{app:LS220-s20}
In Tab.~\ref{tab:FittingParam3} we provide the values of the fitting parameters in Eq.~\eqref{eq:BremFit} for the SN model LS220-s20, employed in Sec.~\ref{sec:TemperatureReconstruction}.
Analogously to the previous cases, the analytic formulas interpolating the time-behavior of the fitting parameters are
\begin{equation}
    \begin{split}
        E^0_{NN}&=\Tilde{E}^0_{NN}\,t_{\mathrm{s}}^{-0.089}\,e^{-0.068\,t_{\mathrm{s}}}\,,\\
        \beta_{NN}&=\Tilde{\beta}_{NN}\,t_{\mathrm{s}}^{0.073}\,e^{-0.037\,t_{\mathrm{s}}}\,,\\
        A_{NN}&=\Tilde{A}_{NN}\,t_{\mathrm{s}}^{0.596}\,e^{-0.693\,t_{\mathrm{s}}}\,,\\
    \end{split}
\end{equation}
in which $\Tilde{E}^0_{NN}=83.55\,\MeV$, $\Tilde{\beta}_{NN}=1.55$ , $\Tilde{A}_{NN}=4.46\times10^{55}\,\MeV^{-1}\,\s^{-1}$.\\

\begin{table}[t!]
\begin{center}

\begin{tabular}{c c c c}
\hline
$t_{\mathrm{pb}}\,[\s]$  & $E^0_{NN}\,[\MeV]$   &$\beta_{NN}$   &$A_{NN}\,[\MeV^{-1}\,\s^{-1}]$ \\
\hline
\hline
1 &  \,\ \,\ 70.19 $ $ $ $ $ $  & \,\ \,\ 1.44 $ $ $ $ $ $  &  \,\ \,\ $4.56\times10^{54}$ $ $ \\
2 &  \,\ \,\  70.39 $ $ $ $ $ $ &  \,\ \,\ 1.42 $ $ $ $ $ $  & \,\ \,\  $4.31\times10^{54}$ $ $ \\
3 &  \,\ \,\  56.91 $ $ $ $ $ $ &  \,\ \,\  1.36 $ $ $ $ $ $  & \,\ \,\  $2.41\times10^{54}$ $ $ \\
4 &  \,\ \,\  58.36 $ $ $ $ $ $ &  \,\ \,\  1.31 $ $ $ $ $ $  &  \,\ \,\ $1.10\times10^{54}$ $ $ \\
5 &  \,\ \,\  47.41 $ $ $ $ $ $ &  \,\ \,\  1.24 $ $ $ $ $ $  &  \,\ \,\ $3.95\times10^{53}$ $ $ \\
6 &  \,\ \,\  35.02 $ $ $ $ $ $ &  \,\ \,\  1.17 $ $ $ $ $ $  &  \,\ \,\ $1.04\times10^{53}$ $ $ \\
7 &  \,\ \,\  23.98 $ $ $ $ $ $ &  \,\ \,\  1.12 $ $ $ $ $ $  &  \,\ \,\ $2.20\times10^{52}$ $ $ \\
8 &  \,\ \,\  16.10 $ $ $ $ $ $ &  \,\ \,\  1.10 $ $ $ $ $ $  &  \,\ \,\ $4.01\times10^{51}$ $ $ \\

\hline
\end{tabular}

\vspace{0.5cm}

\begin{tabular}{c c c c c }
\hline
$t_{\mathrm{pb}}\,[\s]$  & $E^{0}_{\pi N}\,[\MeV]$     &$\beta_{\pi N}$     &$A_{\pi N}\,[\MeV^{-1}\,\s^{-1}]$     &$\omega_c\,[\MeV]$\\
\hline
\hline
1 &  \,\ \,\ 126.43 $ $ $ $ $ $  &  \,\ \,\ 1.20 $ $ $ $ $ $  & \,\ \,\  $2.77\times10^{54}$ $ $ $ $ $ $  & \,\ \,\  103.27$ $\\
2 & \,\ \,\ 94.47 $ $ $ $ $ $  & \,\ \,\ 1.03 $ $ $ $ $ $  & \,\ \,\ $1.24\times10^{54}$ $ $ $ $ $ $  & \,\ \,\ 98.87$ $\\
3 &  \,\ \,\ 56.14 $ $ $ $ $ $  &  \,\ \,\ 0.54 $ $ $ $ $ $  &  \,\ \,\ $9.78\times10^{52}$ $ $ $ $ $ $  & \,\ \,\ 107.00$ $\\
4 & \,\ \,\  37.20 $ $ $ $ $ $  &  \,\ \,\ 0.65 $ $ $ $ $ $  &  \,\ \,\ $2.20\times10^{52}$ $ $ $ $ $ $  & \,\ \,\ 107.06$ $\\
5 &  \,\ \,\ 25.02 $ $ $ $ $ $  &  \,\ \,\ 0.47 $ $ $ $ $ $  & \,\ \,\  $3.63\times10^{51}$ $ $ $ $ $ $  & \,\ \,\ 108.59$ $\\
6 &  \,\ \,\ 15.62 $ $ $ $ $ $  & \,\ \,\  0.40 $ $ $ $ $ $  &  \,\ \,\ $2.53\times10^{50}$ $ $ $ $ $ $  & \,\ \,\ 108.04$ $\\
7 &  \,\ \,\ 9.18 $ $ $ $ $ $  &  \,\ \,\  0.37 $ $ $ $ $ $  &  \,\ \,\ $3.10\times10^{48}$ $ $ $ $ $ $  & \,\ \,\ 108.33$ $\\
8 &  \,\ \,\ 5.64 $ $ $ $ $ $  &  \,\ \,\ 0.37 $ $ $ $ $ $  &  \,\ \,\ $6.64\times10^{45}$ $ $ $ $ $ $  & \,\ \,\ 108.37$ $\\

\hline
\end{tabular}

\end{center}
 \caption{Fitting parameters for $NN$ bremsstrahlung and pion conversion emission spectra in Eq.~\eqref{eq:BremFit} and Eq.~\eqref{eq:PionFit}. These values refer to our benchmark SN model SFHo-s18.8. The time interval considered is $t_{\mathrm{pb}}\in[1,8]\s$ with time steps of $1\,\s$. For the considered case, we have set the ALP-proton coupling to $g_{ap}=5\times10^{-10}$.} 
\label{tab:FittingParam}
\end{table}

\begin{table}[t!]
 \centering
\begin{tabular}{c  c  c  c}
\hline
$t_{\mathrm{pb}}\,[\s]$   & $E^0_{NN}\,[\MeV]$   &$\beta_{NN}$   &$A_{NN}\,[\MeV^{-1}\,\s^{-1}]$\\
\hline
\hline
1 &  \,\ \,\ 86.42 $ $ $ $ $ $  & \,\ \,\ 1.46 $ $ $ $ $ $  & \,\ \,\ $1.93\times10^{55}$ $ $ \\
2 & \,\ \,\  85.28 $ $ $ $ $ $ &  \,\ \,\ 1.45 $ $ $ $ $ $  & \,\ \,\ $2.03\times10^{55}$ $ $ \\
3 & \,\ \,\  82.22 $ $ $ $ $ $ &  \,\ \,\ 1.43$ $ $ $ $ $  & \,\ \,\ $1.58\times10^{55}$ $ $ \\
4 & \,\ \,\  78.10 $ $ $ $ $ $ &  \,\ \,\ 1.39 $ $ $ $ $ $  & \,\ \,\ $1.06\times10^{55}$ $ $ \\
5 & \,\ \,\  72.56 $ $ $ $ $ $ &  \,\ \,\ 1.35 $ $ $ $ $ $  & \,\ \,\ $6.53\times10^{54}$ $ $ \\
6 & \,\ \,\  65.74 $ $ $ $ $ $ &  \,\ \,\ 1.31 $ $ $ $ $ $  & \,\ \,\ $3.59\times10^{54}$ $ $ \\
7 & \,\ \,\  57.54 $ $ $ $ $ $ &  \,\ \,\ 1.26 $ $ $ $ $ $  & \,\ \,\ $1.74\times10^{54}$ $ $ \\
8 & \,\ \,\  48.14 $ $ $ $ $ $ &  \,\ \,\ 1.22 $ $ $ $ $ $  & \,\ \,\ $7.22\times10^{53}$ $ $ \\

\hline
\end{tabular}
 \caption{Fitting parameters for $NN$ bremsstrahlung spectrum in Eq.~\eqref{eq:BremFit}. These values refer to the SFHo-s20 SN model. The considered time interval is $t_{\mathrm{pb}}\in[1,8]\s$ with time steps of $1\,\s$. For the considered case, we have set the ALP-proton coupling to $g_{ap}=5\times10^{-10}$.} 
\label{tab:FittingParam2}
\end{table}

\begin{table}[t!]
 \centering
\begin{tabular}{c  c  c  c}
\hline
$t_{\mathrm{pb}}\,[\s]$   & $E^0_{NN}\,[\MeV]$   &$\beta_{NN}$   &$A_{NN}\,[\MeV^{-1}\,\s^{-1}]$\\
\hline
\hline
1 &  \,\ \,\ 77.46 $ $ $ $ $ $  & \,\ \,\ 1.5 $ $ $ $ $ $  & \,\ \,\ $2.22\times10^{55}$ $ $ \\
2 & \,\ \,\  69.52 $ $ $ $ $ $ &  \,\ \,\ 1.51 $ $ $ $ $ $  & \,\ \,\ $1.71\times10^{55}$ $ $ \\
3 & \,\ \,\  62.60 $ $ $ $ $ $ &  \,\ \,\ 1.51$ $ $ $ $ $  & \,\ \,\ $1.07\times10^{55}$ $ $ \\
4 & \,\ \,\  55.89 $ $ $ $ $ $ &  \,\ \,\ 1.49 $ $ $ $ $ $  & \,\ \,\ $5.97\times10^{54}$ $ $ \\
5 & \,\ \,\  50.39 $ $ $ $ $ $ &  \,\ \,\ 1.47 $ $ $ $ $ $  & \,\ \,\ $3.45\times10^{54}$ $ $ \\
6 & \,\ \,\  46.43 $ $ $ $ $ $ &  \,\ \,\ 1.42 $ $ $ $ $ $  & \,\ \,\ $2.21\times10^{54}$ $ $ \\
7 & \,\ \,\  43.34 $ $ $ $ $ $ &  \,\ \,\ 1.39 $ $ $ $ $ $  & \,\ \,\ $1.47\times10^{54}$ $ $ \\
8 & \,\ \,\  40.37 $ $ $ $ $ $ &  \,\ \,\ 1.35 $ $ $ $ $ $  & \,\ \,\ $9.64\times10^{53}$ $ $ \\

\hline
\end{tabular}
 \caption{Fitting parameters for $NN$ bremsstrahlung spectrum in Eq.~\eqref{eq:BremFit}. These values refer to the LS220-s20 SN model. The considered time interval is $t_{\mathrm{pb}}\in[1,8]\s$ with time steps of $1\,\s$. For the considered case, we have set the ALP-proton coupling to $g_{ap}=5\times10^{-10}$.} 
\label{tab:FittingParam3}
\end{table}

\end{document}